\documentclass[twocolumn]{openjournal}

\usepackage{lipsum}
\usepackage{amsmath}
\usepackage{booktabs}
\usepackage{graphicx}
\usepackage{xspace}
\newcommand{\AbacusSummit}{\texttt{AbacusSummit}\xspace}

\usepackage{xcolor}
\usepackage{textgreek}
\usepackage[utf8]{inputenc}
\usepackage[english]{babel}

\usepackage{hyperref}
\hypersetup{
    unicode, 
    colorlinks=true,
    linkcolor=linkcolor,
    citecolor=linkcolor,
    filecolor=linkcolor,
    urlcolor=linkcolor,
}
\usepackage{color,colortbl}
\definecolor{linkcolor}{rgb}{0.0,0.3,0.5}
\usepackage{tensind}
\tensordelimiter{?}
\DeclareGraphicsExtensions{.bmp,.png,.jpg,.pdf}
\usepackage{verbatim}
\usepackage[normalem]{ulem}
\usepackage{orcidlink}
\usepackage{soul}

\begin{document}
\title{Increasing the Sensitivity of Full-Shape Galaxy Clustering Measurements in Configuration-Space with Three-Point Statistics}

\author{Z. Brown$^{1,*}$\orcidlink{0000-0001-8208-7282} \& L. Samushia$^{1}$\orcidlink{0000-0002-1609-5687}}
\email{$^{*}$zacherybrown@ksu.edu}
\affiliation{$^{1}$ Department of Physics, Kansas State University, 116 Cardwell Hall, Manhattan, KS 66506, USA}

\begin{abstract}

We investigate the cosmological constraining power of a compressed line-of-sight--dependent three-point correlation function (3PCF) estimator on small scales ($\lesssim 80 \ h^{-1}\mathrm{Mpc}$) in configuration space, with a particular focus on emission line galaxies (ELGs) targeted by the Nancy Grace Roman Space Telescope's Galaxy Redshift Survey (GRS), and complementary luminous red galaxy (LRG) samples observed by the Dark Energy Spectroscopic Instrument (DESI). These scales avoid the baryon acoustic oscillation (BAO) feature and are therefore expected to provide information that is largely complementary to standard BAO measurements, while retaining partial overlap with full-shape clustering analyses. Our forecasts are based on \AbacusSummit simulations at $z = 1.1$ and $z = 0.8$, populated with galaxies using halo occupation distribution (HOD) models matched to Roman ELG and DESI LRG samples respectively. The three-point measurements are computed with {\tt TriCo}, a fast configuration-space triangle-counting code developed for this analysis. After marginalizing over uncertainties in the galaxy--halo connection, we find that incorporating the 3PCF yields a substantial improvement over two-point statistics alone, tightening the constraint on $\sigma_8$ by a factor of $\sim$$5$ in our fiducial forecast. This gain arises not from a localized feature or specific scale range, but from the cumulative information content across triangle configurations. Restricting to the monopole of the 3PCF captures only part of this information, with the full line-of-sight--dependent measurement providing an additional factor of $\sim$$2$--$3$ improvement over the monopole. Adding line-of-sight--dependent three-point information substantially increases the constraining power of small-scale configuration-space galaxy clustering.


\end{abstract}

\section{Introduction}
\label{sec:intro}

The spatial distribution of galaxies observed by spectroscopic surveys carries a wealth of cosmological information \citep{cole20052df,parkinson2012wigglez,alam2017clustering,adame2025desi7}. 
In the simplest inflationary models, the primordial density fluctuations are well described by a nearly Gaussian random field
\citep{guth1982fluctuations,bardeen1983spontaneous}. Since a Gaussian random field is fully characterized by its two-point correlation function (2PCF), two-point statistics provide a natural starting point for cosmological analyses of galaxy clustering and are comparatively straightforward to model \citep{keitel2011constrained}. However, the late-time galaxy distribution is no longer Gaussian: gravitational evolution, redshift-space distortions, and galaxy bias generate higher-order correlations that can carry additional cosmological information.

The standard and most robust analyses of large-scale structure (LSS) focus on the baryon acoustic oscillation (BAO) feature and full-shape (FS) modeling of galaxy clustering on linear and quasi-linear scales \citep{blake2011wigglez,bautista2021completed,dumerchat2022baryon,montesano2010new,philcox2022boss,adame2025desi5}. 
BAO measurements target the acoustic scale as a function of redshift, providing precise constraints on cosmological distances, while FS analyses additionally use the broadband shape and redshift-space anisotropy of the clustering signal to constrain growth, matter density, and other cosmological parameters. In parallel, many recent works have explored complementary approaches that leverage non-linear information, including small-scale correlation functions, density-split statistics, and void-based analyses \citep{morawetz2025constraining,radinovic2023euclid,sartori2025imprint,chudaykin2026reanalyzing,mancini2024field,kitanidis2021cross,rocher2023desi}.

The three-point correlation function (3PCF) is a natural extension of two-point statistics and a powerful probe of non-linear structure in galaxy clustering. By capturing correlations among galaxy triplets, it encodes information generated by non-linear gravitational evolution, redshift-space distortions, and galaxy bias that is not fully accessible to the 2PCF alone. 
Consequently, a growing body of work has explored the use of the 3PCF and related Fourier-space bispectrum measurements on both large and small scales to enhance cosmological constraints \citep{gil2015power,gagrani2017information,2018MNRAS.478.4500P,2021MNRAS.505..628S,sugiyama2023new,farina2024modeling,novell2024approximations,novellmasot2025fullshape, 2024MNRAS.531.3326B,chudaykin2026reanalyzing,novellmasot2026cosmological,slepian2025power,brown2025constraining}.

Extracting cosmological information from non-linear scales remains a major theoretical challenge, as perturbation theory–based approaches lose accuracy in this regime.
A common alternative is simulation-based modeling, in which clustering statistics are measured directly from numerical simulations and their dependence on cosmological and galaxy--halo connection parameters is calibrated across suites of simulations. This approach has recently been used to model nonlinear galaxy clustering, density-split statistics, wavelet/scattering summaries, and joint clustering--lensing measurements with {\tt AbacusSummit} and related simulation suites \citep{Yuan_2022,Yuan_2023,cuestalazaro2023sunbirdsimulationbasedmodelfullshape,Valogiannis_2024,lange2025cosmologicalconstraintsfullscaleclustering,dumerchat2026emulatinggalaxypeculiarvelocity,hahn2023forward,lemos2024field, Burger_2024}. It enables predictions beyond the reach of simple analytic approximations, provided that the simulations span the relevant cosmological and galaxy--halo parameter space.

In this work, we study the potential cosmological constraints from an analysis of the 3PCF on scales smaller than the BAO feature ($\lesssim 80\,h^{-1}\mathrm{Mpc}$).  By restricting scales to only separations below the BAO peak, the analysis avoids directly reusing the acoustic feature which has been detected in the large scale 3PCF \citep{Slepian_2017, kamalinejad2026detectionbaryonacousticoscillation}. We therefore expect the resulting constraints to be weakly correlated with BAO-only distance measurements, although a full joint covariance with BAO analyses remains to be quantified. We apply this framework to two representative samples:  Roman Space Telescope emission-line galaxy sample at $z \approx 1.1$ and DESI luminous red galaxies (LRGs) at $z \approx 0.8$.

The Roman Space Telescope will carry out the High Latitude Spectroscopic Survey (HLSS), a wide-field slitless spectroscopic survey designed to map large-scale structure over thousands of square degrees at near-infrared wavelengths \citep{2019arXiv190205569A,2021MNRAS.507.1746E,2022ApJ...928....1W}. The HLSS will target emission-line galaxies, primarily through H$\alpha$ and [OIII] emission, over a redshift range where Roman combines large survey volume with high source density \citep{2019MNRAS.490.3667Z,2021MNRAS.501.3490Z,2021arXiv210912216Z}. This makes the Roman ELG sample particularly well suited for small-scale clustering analyses beyond the standard two-point BAO program. In this work, we focus on a representative Roman-like ELG sample at $z\simeq 1.1$, where the expected number density and survey volume provide strong statistical sensitivity to the three-point correlation function. To test whether the improvement from the 3PCF is specific to this favorable high-density sample, or whether it extends to other survey regimes, we also apply the same methodology to DESI LRGs at lower redshift and lower number density.

DESI is an ongoing spectroscopic survey designed to measure the expansion history and growth of structure using several galaxy and quasar tracers over a large fraction of the sky \citep{levi2013desi,aghamousa2016desi1,aghamousa2016desi2}. In this work, we focus on the LRG sample as a complementary test case. DESI LRGs probe a large cosmological volume at intermediate redshift, while occupying a regime in which non-linear clustering remains significant and the galaxy number density is still sufficient for small-scale higher-order clustering measurements. Compared with the lower-redshift BGS sample, the LRG sample provides access to a substantially larger volume; compared with higher-redshift tracers such as ELGs and QSOs, it retains stronger non-linear clustering signal on the scales considered here \citep{prieto2020preliminary,ruiz2020preliminary,zhou2020preliminary,yeche2020preliminary,lan2023desi,alexander2023desi,cooper2023overview,hahn2023desi,zhou2023target,raichoor2023target,chaussidon2023target}.

The primary methodological contribution of this paper is a compressed LOS-dependent 3PCF basis that preserves anisotropic information relevant for redshift-space clustering while keeping the data vector small enough for stable covariance estimation. The basis is complementary to other existing 3PCF data reduction schemes \citep{slepian2015computing, sugiyama2019triposh}. We quantify, using \AbacusSummit simulations and HOD marginalization, how much this basis improves cosmological forecasts relative to conventional 2PCF multipoles and to a LOS-averaged 3PCF measurement.

Because the \AbacusSummit simulations contain dark matter halos rather than galaxies, uncertainties in the galaxy--halo connection must be modeled explicitly. We populate halos using the standard HOD framework implemented in {\tt AbacusHOD} \citep{yuan2022abacushod}. In the forecasts below, the HOD parameters are treated as nuisance parameters and marginalized over, allowing us to quantify the cosmological information in the 2PCF and 3PCF after accounting for galaxy--halo uncertainties. While the HOD parameter space can be explored continuously within this framework, the cosmological parameter space is sampled only at discrete points defined by the \AbacusSummit simulation grid. Rather than constructing a full emulator, we adopt a linear expansion around a fiducial cosmology. We validate this approximation using additional \AbacusSummit cosmologies.

The main novelty of this work is the combination of a compact LOS-dependent configuration-space 3PCF estimator, fast triangle counting with TriCo, and simulation-based HOD-marginalized forecasts for Roman-like and DESI-like galaxy samples.

Our paper is organized as follows. Section~\ref{sec:sim_details} describes the \AbacusSummit simulations and HOD galaxy mocks used in this work. Section~\ref{sec:clustering_stats} defines the 2PCF and LOS-dependent 3PCF measurements and describes the construction of the data vector. Section~\ref{sec:model_predictions} presents the linear response model, and Section~\ref{sec:covariance} the covariance treatment. Section~\ref{sec:cosmo_forecasts} gives the cosmological forecasts and quantifies the information gained by including the 3PCF. Section~\ref{sec:robustness} validates the linear approximation using additional simulations. Section~\ref{sec:scale_dep} shows how our choice of scales affects the resultant constraints. We discuss the implications and limitations of the results in Section~\ref{sec:discussion_conclusions}.

\section{Simulation Details}
\label{sec:sim_details}

The {\tt AbacusSummit} simulations consist of periodic $(2\,h^{-1}{\rm Gpc})^3$ volumes evolved with $6912^3$ dark matter particles, corresponding to a particle mass of approximately $2\times 10^9,h^{-1}M_\odot$ \citep{maksimova2021abacussummit,garrison2021abacus}. In this work, we use snapshots at redshifts $z=1.1$ and $z=0.8$, which we use to construct Roman-like ELG and DESI-like LRG samples, respectively. For the fiducial cosmology, we adopt the Planck 2018--based AbacusSummit cosmology {\tt c000} \citep{2020A&A...641A...6P}.

To model the response of galaxy clustering to cosmological parameters, we additionally use the AbacusSummit cosmologies {\tt c100}, {\tt c102}, {\tt c104}, {\tt c106}, {\tt c112}, {\tt c114}, and {\tt c121}. These simulations vary the dimensionless Hubble parameter $h$, the cold dark matter density $\omega_{\rm cdm}$, the scalar spectral index $n_s$, the running $\alpha_s=d n_s/d\ln k$, the fluctuation amplitude $\sigma_8$, the effective number of ultra-relativistic species $N_{\rm ur}$, and the dark-energy equation-of-state parameter $w_0$.

\begin{equation}
\label{eq:cosmo_pois}
\theta^* = [h,\omega_{\rm cdm},n_s,\alpha_s,\sigma_8,N_{\rm ur},w_0] \ ,
\end{equation}
In the fiducial massive-neutrino cosmology {\tt c000}, AbacusSummit uses one massive neutrino species with $\omega_{\rm ncdm}=0.00064420$. We emphasize that $N_{\rm ur}$ controls the density of additional massless (ultra-relativistic) species and therefore corresponds to a variation in the effective number of relativistic species $N_{\rm eff}$ at fixed neutrino mass. Throughout, we hold the single massive-neutrino species fixed at $\omega_{\rm ncdm}=0.00064420$, i.e.\ $\sum m_\nu \simeq 0.06\,{\rm eV}$. Our forecasts thus probe the $N_{\rm eff}$ direction rather than the absolute neutrino-mass scale, and the constraints we quote on $N_{\rm ur}$ should not be read as neutrino-mass constraints.

In our analysis, we parameterize the matter density using $\omega_{\rm cdm}$ rather than $\Omega_m$, since $\omega_{\rm cdm}$ more directly controls the physical cold-dark-matter density relevant for the clustering calculation, while $\Omega_m$ also depends on the Hubble parameter and is closely tied to late-time distance measurements. For the AbacusSummit massive-neutrino cosmologies, the total matter density is related to the physical densities by
\begin{equation}
\label{eq:matter_dens}
\Omega_m = \frac{\omega_{\rm cdm}+\omega_b+\omega_{\rm ncdm}}{h^2} \ .
\end{equation}
The baryon density is fixed to $\omega_b\equiv\Omega_b h^2=0.02237$, consistent with the Planck 2018 best-fit value, since it is tightly constrained by CMB measurements at the sub-percent level \citep{2020A&A...641A...6P}.

We populate halos with galaxies using the standard HOD prescription implemented in the {\tt AbacusHOD} pipeline \citep{yuan2022abacushod}. The full HOD model contains five parameters,
\begin{equation}
\theta^{\rm HOD}_{\rm full} =
[M_{\rm cut}, \sigma, \kappa, M_1, \alpha],
\end{equation}
which control the mean number of central and satellite galaxies assigned to halos as a function of halo mass. In the parameter inference below, we vary $\log M_{\rm cut}$, $\sigma$, $\log M_1$, and $\alpha$. Since our measurements are restricted to separations above $2.5,h^{-1}{\rm Mpc}$, we find that varying $\kappa$ within reasonable prior bounds has a negligible impact on our results. We therefore fix $\kappa=0.55$.

For a halo of mass $M$, the mean number of central galaxies is
\begin{equation}
\label{eq:hod_defs_cent}
\bar{n}_{\rm cent}(M) =
\frac{1}{2}{\rm erfc}
\left[
\frac{\log(M_{\rm cut}/M)}{\sqrt{2}\sigma}
\right],
\end{equation}
where ${\rm erfc}$ is the complementary error function. The mean number of satellite galaxies is
\begin{equation}
\label{eq:hod_defs_sat}
\bar{n}_{\rm sat}(M) =
\left[
\frac{M-\kappa M_{\rm cut}}{M_1}
\right]^\alpha
\bar{n}_{\rm cent}(M),
\end{equation}
with $\bar n_{\rm sat}=0$ for $M\leq \kappa M_{\rm cut}$.

Central galaxies are assigned the positions and velocities of halo centers, while satellite galaxies are randomly assigned to dark matter particles within each halo, thereby inheriting the spatial and velocity distribution of the halo particles. In this work, we do not include central or satellite velocity bias. We have verified that reasonable variations of the velocity-bias parameters have a subdominant impact on the scales considered here, and therefore fix them in the fiducial analysis.

For the Roman-like ELG sample, we adopt the target number densities from \citet{spezzati2026forecastingneutrinomassconstraints} and tune the HOD parameters to reproduce those densities. For the DESI-like LRG sample, we choose HOD parameters consistent with recent DESI LRG HOD measurements and mock-catalog studies \citep{yuan2023desionepercentsurveyexploring,Mena_Fern_ndez_2025}. Both fiducial galaxy samples have satellite fractions of approximately $15\%$, with the remaining galaxies assigned as centrals.

We adopt this baseline HOD form for both samples. For the ELG sample this is a deliberate simplification: star-forming ELGs are known to avoid the most massive, quenched halos, so their central occupation is not well described by the monotonically rising error-function form of Eq.~\eqref{eq:hod_defs_cent}, and dedicated DESI analyses model them with a high-mass-quenched (modified-Gaussian) central term instead \citep{rocher2023desi}. We do not adopt such a form here for two reasons. First, our measurements are restricted to $r>2.5\,h^{-1}{\rm Mpc}$ and exclude the deeply non-linear one-halo regime, where the detailed shape of the central and satellite occupation has the largest effect on the clustering. Second, this work is a forecast aimed at the \emph{relative} information gain from the 3PCF rather than at absolute parameter values, so the leading requirement on the mock is that it reproduce the correct number density and approximate large-scale bias and satellite fraction, which the tuned baseline HOD achieves by construction. We therefore expect the choice of ELG central parameterization to have a subdominant impact on our forecast conclusions, and defer a quenching-aware ELG HOD to analyses of real data.

In addition to the response to cosmological parameters, our Fisher forecasts require the response of the clustering measurements to the HOD parameters. We estimate these derivatives by generating additional galaxy catalogs at the fiducial cosmology {\tt c000}, varying one HOD parameter at a time while keeping the other HOD and cosmological parameters fixed to their fiducial values. The resulting HOD-variation simulations are labeled {\tt c901}, {\tt c902}, {\tt c904}, and {\tt c905}, corresponding to variations in the four free HOD parameters.

\begin{table}[t]
    \centering
    \renewcommand{\arraystretch}{1.5}
    \begin{tabular}{ c c c }  
        Parameter & DESI LRG Fid. & Roman ELG Fid. \\
        \hline \hline  
        $h$ & 0.6736 & 0.6736 \\
        $\omega_{\mathrm{cdm}}$ & 0.12 & 0.12\\
        $n_s$ & 0.9649 & 0.9649\\
        $\alpha_s$ & 0.0 & 0.0\\
        $\sigma_8$ & 0.807952 & 0.807952\\
        $N_{\mathrm{ur}}$ & 2.0328 & 2.0328\\
        $w_0$ & -1.0 & -1.0\\
        $\log M_{\mathrm{cut}}$ & 12.78 & 12.32 \\
        $\sigma$ & 0.21 & 0.25 \\
        $\log M_1$ & 13.88& 13.45 \\
        $\alpha$ & 1.07 & 1.05 \\
        $\kappa$ & 0.55 (fixed) & 0.55 (fixed)\\
        \hline \hline  
    \end{tabular}
    \caption{The values of the fiducial cosmological and HOD parameters in this study.}
    \label{table:fid_pars}
\end{table}

\begin{figure}[t]
    \centering
    \includegraphics[width=0.45\textwidth]{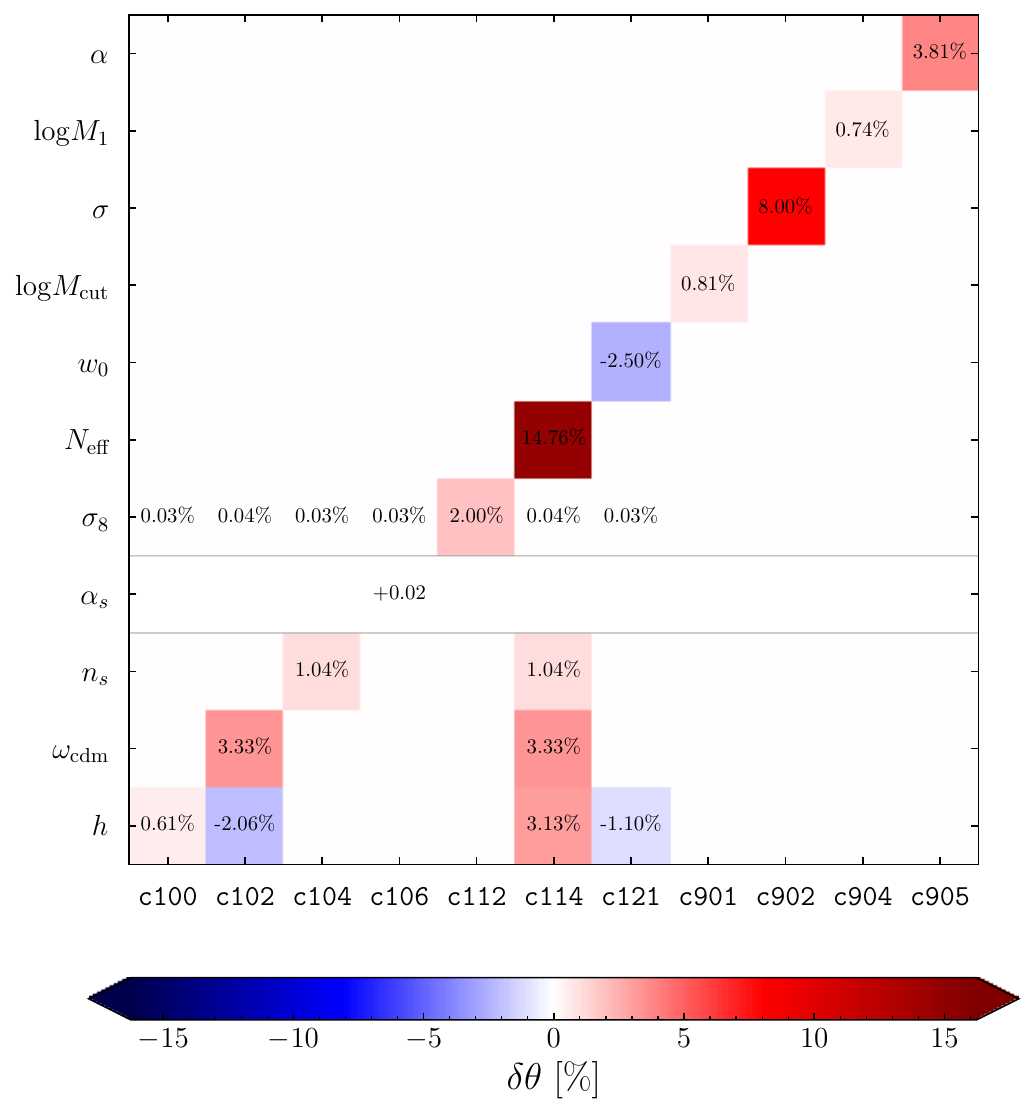}
    \caption{The percent variation of each parameter in our model with respect to the fiducial case for 11 different simulations used in our forecasts (color scale). All parameters are shown as percentage deviations except for $\alpha_s$, where the true value is shown for cosmology {\tt c106}.}
    \label{fig:par_grid}
\end{figure}

The fiducial values of the cosmological and HOD parameters are given in Tab.~\ref{table:fid_pars} for both DESI LRGs and Roman ELGs. Figure~\ref{fig:par_grid} shows the variation of each parameter across the simulations used in this study for the Roman ELGs. All parameter variations are expressed as percentages relative to their fiducial values, except for $\alpha_s$, which is shown in absolute units. The fiducial value of $\alpha_s$ is 0.0 for all simulations except {\tt c106}, where it is set to $\alpha_s = 0.02$.

The offset simulations corresponding to variations in the HOD parameters, as well as in $\sigma_8$ and $\alpha_s$, are constructed by varying each parameter individually while keeping all other parameters fixed to their fiducial values. In contrast, variations in $N_{\mathrm{ur}}$, $n_s$, $\omega_{\mathrm{cdm}}$, and $h$ are not implemented as independent one-dimensional steps. Instead, these parameters are varied in a correlated manner to span a four-dimensional subspace of cosmological parameter space while keeping $\sigma_8$ fixed. This design choice follows the \AbacusSummit simulation strategy.

\section{Clustering Statistics}
\label{sec:clustering_stats}

As summary statistics, we consider the galaxy 2PCF and 3PCF. The 2PCF is computed using the {\tt Corrfunc} package \citep{sinha2020corrfunc}. We restrict our analysis to scales below the baryon acoustic oscillation (BAO) feature so that our measurements are effectively independent of BAO information. Galaxy pairs are binned by their separation $s$ into 16 bins of width $5\,h^{-1}\mathrm{Mpc}$, spanning $2.5$ to $82.5\,h^{-1}\mathrm{Mpc}$ (corresponding to the lower edge of the first bin and the upper edge of the last bin). The bin centers are located at $5, 10, 15, \dots\,h^{-1}\mathrm{Mpc}$.

Pairs are additionally binned by the cosine of the angle between their separation vector and the line-of-sight (LOS), $\mu = \cos\theta$. The resulting pair-count histogram, $DD(s,\mu)$, is normalized and projected onto Legendre polynomials to obtain the multipoles of the 2PCF. The full two-dimensional correlation function is defined as
\begin{equation}
    \label{eq:2PCF_2d}
    \xi(s,\mu)  = \frac{DD(s,\mu)}{RR(s,\mu)} - 1 \ ,
\end{equation}
where $RR(s,\mu)$ is the expected number of pairs for a random distribution. For a periodic box containing $N_{\mathrm{gal}}$ galaxies in a volume $V$, this is given by
\begin{equation}
    \label{eq:rr_norm}
    RR(s,\mu)  =  \frac{4 \pi \left( s_{\mathrm{max}}^3 - s_{\mathrm{min}}^3 \right) N_{\mathrm{gal}}^2}{3 V N_{\mathrm{bins}}^{\mu}} \ ,
\end{equation}
where $s_{\mathrm{max}}$ and $s_{\mathrm{min}}$ define the bin edges, and $N_{\mathrm{bins}}^{\mu}$ is the number of $\mu$ bins. We adopt 40 bins in $\mu$ over the range $0 \leq \mu \leq 0.9$, thereby excluding pairs that are most closely aligned with the LOS. The normalization is chosen to match the pair-count convention used by {\tt Corrfunc} and the restricted range \(0\leq\mu\leq0.9\). The $\ell^{\mathrm{th}}$ multipole of the 2PCF is then computed as
\begin{equation}
    \label{eq:2PCF_expansion}
    \xi_{\ell}(s) = \sum_{\mu - \mathrm{bins}} \xi(s,\mu)\, P_{\ell}(\mu)\, \Delta \mu \ .
\end{equation}
In this analysis, we consider the monopole ($\ell = 0$) and quadrupole ($\ell = 2$).

The exclusion of large-$\mu$ pairs is motivated by the anticipated application of this method to survey data. In spectroscopic surveys, pairs of galaxies aligned along the LOS are more difficult to observe due to fiber collisions and projection effects, leading to an artificial suppression of high-$\mu$ pair counts. While this effect is absent in simulations, applying a simulation-based model to survey data without accounting for it could introduce systematic biases. Excluding high-$\mu$ pairs mitigates this potential source of systematic error \citep{bianchi2025desi_fiber_dr1,lasker2025desi_pip,bianchi2018fiber_assignment}. 

\begin{figure*}[!t]
    \centering
    \includegraphics[width=\textwidth]{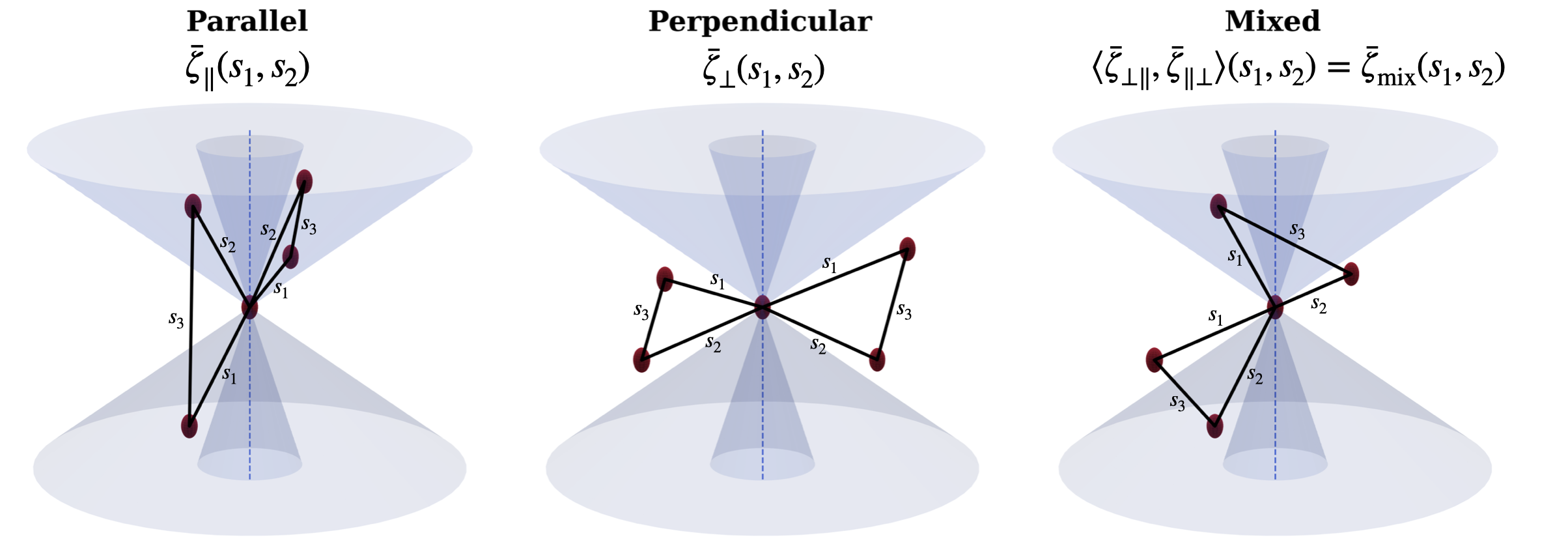}
    \caption{A schematic of the 3PCF decomposition employed by the {\tt TriCo} algorithm. The 3PCF is binned according to the length of two sides of the triangle, $s_1$ and $s_2$, as well as the cosine of the angles those sides make with respect to the line-of-sight, $\mu_1 = \cos \theta_1$ and $\mu_2 = \cos \theta_2$. In the figure, the line-of-sight is shown as a blue dashed line. The shaded cones represent the boundaries of the $\mu$-bins, and two triangle examples are shown for each case.}
    \label{fig:trico_3pt}
\end{figure*}


Additional cosmological information is encoded in higher-order clustering statistics beyond the two-point function. We therefore consider the three-point correlation function (3PCF) and introduce a novel estimator designed to efficiently capture its line-of-sight (LOS) dependence. This estimator is implemented in the {\tt TriCo} algorithm\footnote{\url{https://github.com/ladosamushia/TriCo}}.

Ideally, one would characterize each triangle by the lengths of all three sides together with its orientation relative to the line-of-sight (LOS), for example through two independent angles. Such a description fully specifies the configuration of a galaxy triplet, but in practice leads to a very high-dimensional data vector, making covariance estimation and likelihood analyses computationally challenging and potentially numerically unstable \citep{umeh2021optimal}. For this reason, modern 3PCF estimators adopt compressed representations of the full configuration space.

A widely used approach (e.g., \citealt{slepian2015computing}) parametrizes triangles by two side lengths, $s_1$ and $s_2$, and performs a Legendre decomposition with respect to the opening angle between them. This provides an efficient basis that captures much of the relevant information while significantly reducing dimensionality. In this work, we adopt a similar philosophy of data compression, but focus on retaining more explicit control over the LOS dependence of the signal. In particular, we aim to distinguish between triangle configurations that are preferentially aligned with the LOS and those that lie predominantly transverse to it.

We define a LOS-dependent decomposition of the 3PCF that retains explicit information about the orientation of triangle configurations with respect to the LOS. For each galaxy triplet, we adopt the standard definition of triangle sides $s_1$ and $s_2$, and additionally introduce two angles, $\theta_1$ and $\theta_2$, between each side and the LOS. We then bin triangles according to $\mu_1 = \cos\theta_1$ and $\mu_2 = \cos\theta_2$.

In this work, each $\mu_i$ ($i=1,2$) is classified as either approximately LOS-aligned 
($\parallel$: $0.45 < \mu_i < 0.90$) or approximately transverse 
($\perp$: $0 < \mu_i < 0.45$). Configurations with $\mu_i > 0.90$ are excluded to avoid the most strongly LOS-aligned pairs, which are expected to be most sensitive to fiber-assignment and projection effects in survey applications.

In principle, the configurations $(\parallel,\perp)$ and $(\perp,\parallel)$ are not identical, as they correspond to different assignments of $s_1$ and $s_2$. However, we average over these two cases, as the difference between their expectation values carries little additional information for our purposes. We define this case as the ``mixed'', where $\bar{\zeta}_{\rm mix}(s_1,s_2) \equiv \frac{1}{2} \left[ \bar{\zeta}_{\parallel,\perp}(s_1,s_2) + \bar{\zeta}_{\perp,\parallel}(s_1,s_2) \right]$. This choice reduces the dimensionality of the data vector while preserving the dominant LOS-dependent signal.

The resulting LOS-binned 3PCF, denoted $\bar{\zeta}_{j}(s_1,s_2)$ with $j \in \{\parallel,{\rm mix},\perp\}$, is constructed from data and random triplet counts as

\begin{equation}
    \label{eq:3PCF_def}
    \bar{\zeta}_{j}(s_1,s_2) = \frac{DDD_{j}(s_1,s_2)}{RRR_{j}(s_1,s_2)} - 1 \ .
\end{equation}
The random counts $RRR_{j}(s_1,s_2)$ are computed using a dense uniform distribution of particles. While these counts can in principle be evaluated analytically for a periodic box, the numerical computation is sufficiently fast that we opt for a direct estimation \citep{PearsonSamushia2019}. In this work, \(\bar{\zeta}\) denotes the excess triplet count in the specified configuration bin. For a periodic box this quantity can be estimated directly as \(DDD/RRR-1\). It differs from the connected 3PCF by the contribution of the three two-point terms; throughout this paper we use this compressed triplet statistic as the observable entering the forecasts.

This results in three distinct classes of triangle configurations: those with both selected sides more LOS-aligned, those with both selected sides more transverse, and mixed configurations.

In our analysis, the observable, $\tilde{O}$, is constructed from the 2PCF monopole and quadrupole, together with the three LOS-binned components of the 3PCF. Each of the 3PCF measurements is flattened into a one-dimensional data vector. Explicitly,
\begin{equation}
    \label{eq:data_vector}
    \tilde{O} = \left[ \xi_0, \xi_2, \bar{\zeta}_{\parallel}, \bar{\zeta}_{\rm mix}, \bar{\zeta}_{\perp} \right].
\end{equation}
The 2PCF multipoles, $\xi_{\ell=0}$ and $\xi_{\ell=2}$, are ordered by increasing separation $s$. Each 3PCF component is ordered first by increasing $s_1$ and then by increasing $s_2$. For the binning scheme described in this section, $\tilde{O}$ contains 440 elements. We use the same radial binning for \(s_1\) and \(s_2\) as for the 2PCF and retain only ordered pairs with \(s_1 \geq s_2\), so that each triangle configuration is counted once in the compressed data vector.

    \begin{figure}[!t]
        \centering
    	\includegraphics[width=0.45\textwidth]{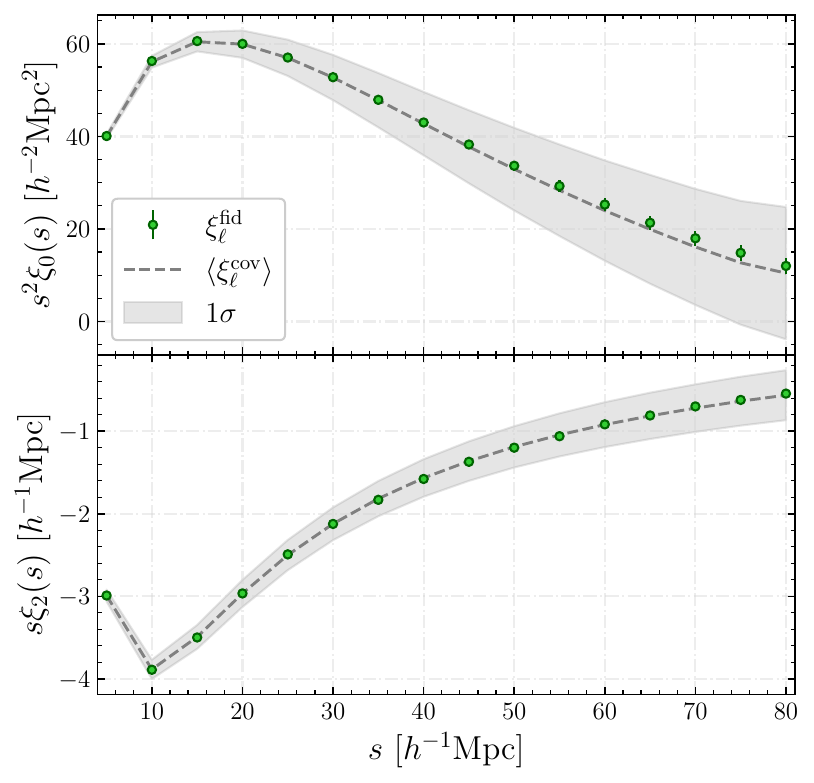}
    	\caption{The 2PCF, $\xi_{\ell}(s)$ of the Roman ELG fiducial case (green) and of the simulations used to construct the covariance matrix (grey). The grey shaded band corresponds to the 1$\sigma$ errors of the small (500 $h^{-1}$ Mpc)$^3$ cubes. The error bars on the green markers correspond to the 1$\sigma$, scaled according to the (2000 $h^{-1}$ Mpc)$^3$ volume of the large box. The upper panel shows the monopole ($\ell=0$), while the lower panel shows the quadrupole ($\ell=2$).}
        \label{fig:xi_fid}
    \end{figure}

    \begin{figure}[!t]
        \centering
    	\includegraphics[width=0.45\textwidth]{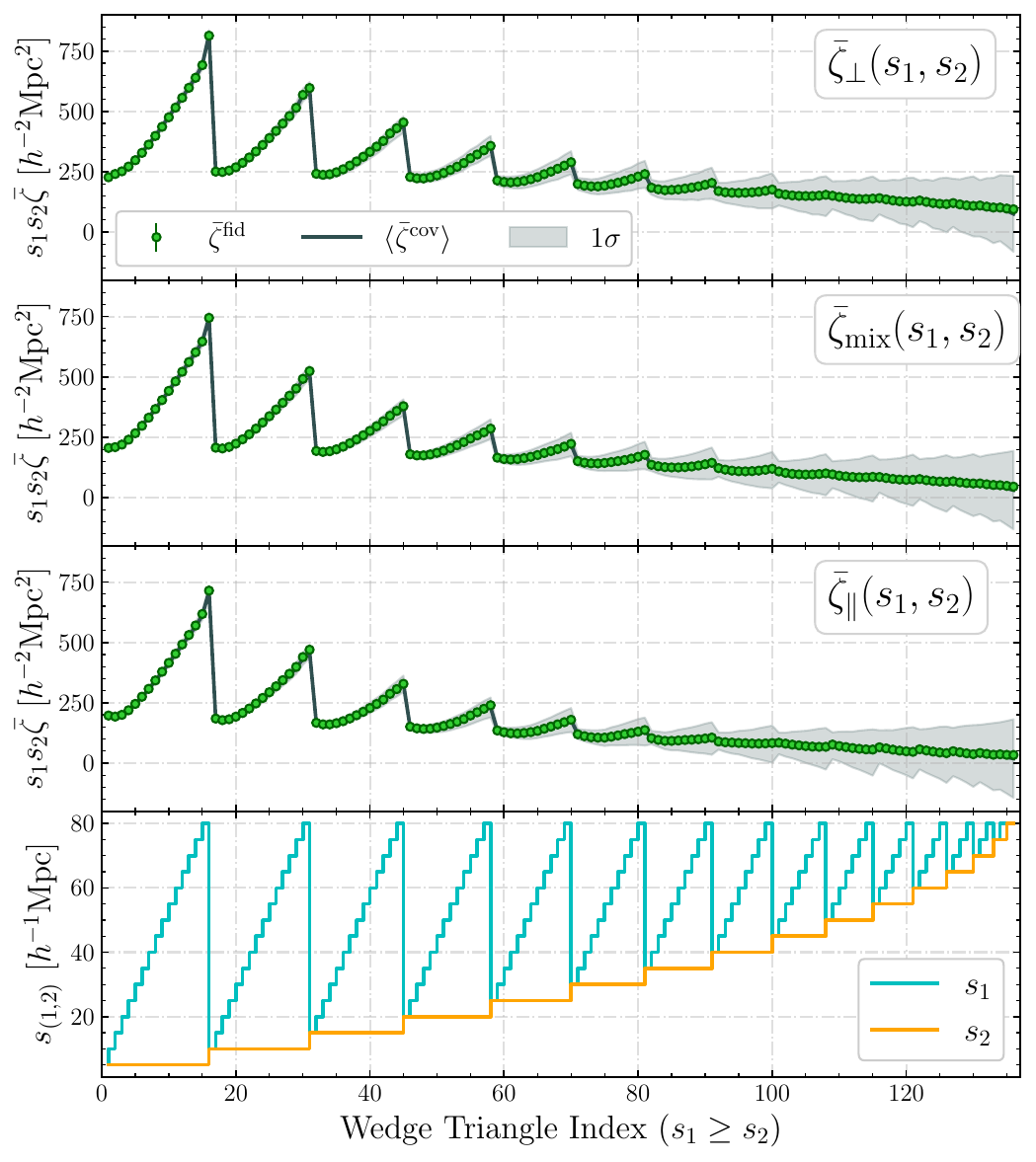}
    	\caption{The 3PCF, $\bar{\zeta}(s_1,s_2)$ of the Roman ELG fiducial case (green) and of the simulations used to construct the covariance matrix (grey). The grey shaded band corresponds to the 1$\sigma$ errors of the small (500 $h^{-1}$ Mpc)$^3$ cubes. The error bars on the green markers correspond to the 1$\sigma$, scaled according to the (2000 $h^{-1}$ Mpc)$^3$ volume of the large box. The lower panel shows the length of each triangle side $s_1$ and $s_2$.}
        \label{fig:zeta_fid}
    \end{figure}

The fiducial 2PCF and 3PCF measurements are shown in Figs.~\ref{fig:xi_fid} and~\ref{fig:zeta_fid}.

\section{Model Predictions}
\label{sec:model_predictions}

We model the data vector \(\widetilde{O}\) using a linear expansion about the fiducial cosmological and HOD parameter values specified in Table~\ref{table:fid_pars}. The required derivatives are estimated from the set of offset simulations shown in Fig.~\ref{fig:par_grid}, which include both cosmological and HOD parameter variations.

Let \(\theta^{\rm fid}\) denote the fiducial value of the combined 11-dimensional parameter vector,
\begin{equation}
\theta =
[h,\omega_{\rm cdm},n_s,\alpha_s,\sigma_8,N_{\rm eff},w_0,
\log M_{\rm cut},\sigma,\log M_1,\alpha] ,
\end{equation}
and define the parameter displacement
\begin{equation}
    \delta\theta = \theta-\theta^{\rm fid}.
\end{equation}
The fiducial data vector, \(\widetilde{O}^{\rm fid}\), contains the 2PCF and 3PCF measurements evaluated at the fiducial parameter values and ordered as in Eq.~\ref{eq:data_vector}.

To estimate the response of the data vector to parameter variations, we define a matrix \(P\) whose element \(P_{ij}\) is the displacement of the \(j\)-th parameter in the \(i\)-th offset simulation. We also define \(\Delta\) as the corresponding change in the data vector,
\begin{equation}
    \Delta_{ij}
    =
    \widetilde{O}_{j}(\theta_i^{\rm off})
    -
    \widetilde{O}^{\rm fid}_{j},
\end{equation}
where \(\theta_i^{\rm off}\) denotes the parameter vector of the \(i\)-th offset simulation. The Jacobian of the data vector with respect to the model parameters is then estimated as
\begin{equation}
    J = P^{-1}\Delta .
\end{equation}
This form accounts for the fact that the parameter displacements in the simulation grid are not all aligned with individual coordinate directions.

The model prediction at parameter values \(\theta\) is therefore
\begin{equation}
    \label{eq:model_linear}
    \widetilde{O}(\theta)
    =
    \widetilde{O}^{\rm fid}
    +
    \delta\theta\,J
    =
    \widetilde{O}^{\rm fid}
    +
    \delta\theta \cdot P^{-1}\cdot \Delta .
\end{equation}
This is the first-order Taylor expansion of the data vector around the fiducial model.

\begin{figure}[!t]
    \centering
    \includegraphics[width=0.4\textwidth]{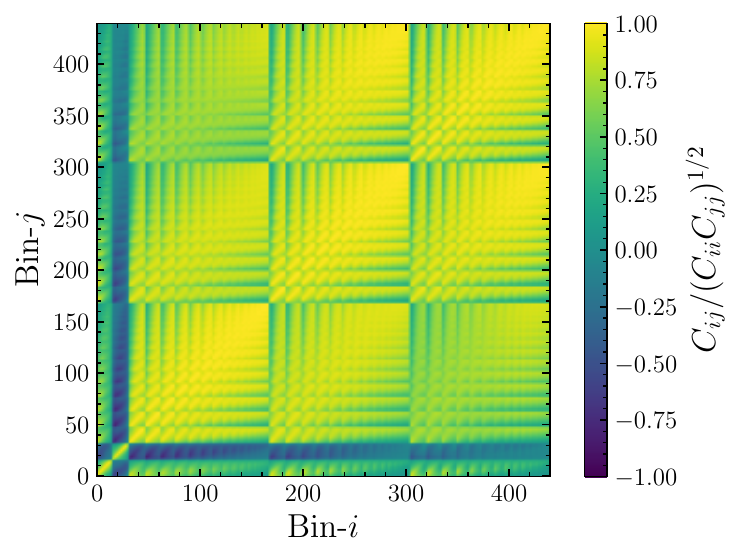}
    \caption{Normalized covariance matrix, $C_{ij}/\sqrt{C_{ii}C_{jj}}$, estimated from 1000 {\tt Abacus} simulations corresponding to the Roman ELG density. The matrix elements are ordered according to Eq.~\ref{eq:data_vector}. The covariance computed from $(500\,h^{-1}\mathrm{Mpc})^3$ boxes is rescaled to match the volume of the fiducial simulations.}
    \label{fig:cov_mat}
\end{figure}

\section{Covariance Matrix}
\label{sec:covariance}

The covariance entering our forecasts has two conceptually distinct contributions. The first is the covariance of the clustering measurements for a finite survey volume, while the second is the stochastic uncertainty in the simulation-based model used to predict the data vector. We write the total covariance schematically as
\begin{equation}
    C^{\rm tot} = C^{\rm meas} + C^{\rm mod} ,
\end{equation}
assuming that the measurement and model uncertainties are uncorrelated.

For an analysis of real survey data, \(C^{\rm meas}\) would include the effects of survey geometry, selection functions, fiber assignment, non-uniform completeness, and other observational systematics. In this work, however, our goal is not to produce a survey-specific DESI or Roman forecast. Instead, we use periodic-box simulations to estimate the relative gain in constraining power obtained by adding three-point information to the two-point function at fixed effective volume. The covariance used below should therefore be interpreted as an idealized effective-volume covariance rather than as the covariance of a particular survey sample.

We estimate this covariance from 1000 independent Abacus simulations at the fiducial cosmology. Each small box has volume \((500\,h^{-1}{\rm Mpc})^3\). The sample covariance of the data vector \(\widetilde O\) measured from these boxes is rescaled by the ratio of volumes to obtain the covariance corresponding to a full \((2\,h^{-1}{\rm Gpc})^3\) Abacus box,
\begin{equation}
    C^{\rm Ab} = \frac{1}{64} C^{500},
\end{equation}
where \(C^{500}\) denotes the covariance measured from the smaller boxes. This rescaling assumes the dominant covariance terms scale approximately inversely with volume over the range of scales considered here. The resulting normalized covariance matrix is shown in Fig.~\ref{fig:cov_mat}.

This covariance estimate is approximate in several respects. Since it is constructed from periodic boxes, it does not include survey-window effects or observational systematics. In addition, rescaling the covariance from smaller boxes neglects possible contributions from modes larger than the small-box volume, including super-sample covariance. Finally, the inverse covariance is affected by the finite number of realizations used to estimate \(C^{500}\). These effects are important for a production analysis of survey data, but they are expected to have a subdominant impact on the relative comparison between 2PCF-only and 2PCF+3PCF forecasts performed here. We therefore use \(C^{\rm Ab}\) as an effective covariance for assessing the information gain from the 3PCF.

The model prediction in Eq.~\ref{eq:model_linear} is itself derived from simulations and therefore also carries stochastic uncertainty. The fiducial data vector \(\widetilde O^{\rm fid}\) is averaged over 25 realizations, giving a covariance contribution of approximately \(C^{\rm Ab}/25\), which is small compared with the other terms and is neglected. The derivative matrix, however, is estimated from offset simulations with finite volume. If the stochastic fluctuations in the offset simulations are treated as independent of those in the fiducial mean, then propagation of this uncertainty through the linear model gives
\begin{equation}
\label{eq:mod_covariance}
C^{\rm mod}(\theta)
=
\left[
\delta\theta \cdot P^{-1} \cdot (P^{-1})^{T} \cdot \delta\theta^{T}
\right]
C^{\rm Ab}.
\end{equation}
This expression vanishes at the fiducial point and increases with distance from the fiducial model. At a parameter displacement corresponding to one of the offset simulations used to estimate the derivatives, the prefactor is unity, so that \(C^{\rm mod}=C^{\rm Ab}\).

In principle, one could include this parameter-dependent covariance directly in the likelihood. Doing so would require using the full Gaussian likelihood, including the \(\log |C^{\rm tot}(\theta)|\) normalization term. However, because our mock data vector is defined to be the fiducial mean, the best-fit residual vanishes by construction. In this setup, a parameter-dependent model covariance can obscure the interpretation of the forecast by changing the effective width of the likelihood as one moves away from the fiducial point. We therefore adopt a fixed effective covariance for the forecasts. Specifically, we use
\begin{equation}
    C^{\rm tot} = C^{\rm Ab} + C^{\rm mod}_{\rm eff} \ ,
\end{equation}
with
\begin{equation}
    C^{\rm mod}_{\rm eff}=C^{\rm Ab} \ .
    \label{eq:cov_equality}
\end{equation}
This choice corresponds to the model uncertainty expected at a characteristic one-step displacement in the simulation grid. It is conservative near the fiducial point, where the formal model covariance is smaller, and provides a simple common covariance with which to compare different choices of data vector.

We have verified that correlations between the fiducial mean and the derivative estimates are weak on the scales used in this analysis, particularly for the 3PCF, and therefore neglect them in the covariance propagation. The covariance matrix is invertible and sufficiently well conditioned for the adopted data vector. In the final analysis, we will quantify the stability of the forecasts by varying the effective model covariance and by testing the sensitivity of the results to the smallest-eigenvalue modes of the covariance matrix.

From Eq.~\ref{eq:cov_equality}, we define the total covariance as twice $C^{\rm Ab}$, with 
\begin{equation}
    C^{\rm tot} = 2 C^{\rm Ab} \ .
    \label{eq:cov_equality}
\end{equation}
Upon its inversion, we apply the Hartlap correction \citep{hartlap2007your} to our covariance matrix. When using $N_m$ mock catalogs to estimate the covariance of a data vector consisting of $N_{\rm data}$ entries, the inverse covariance should be scaled according to 
\begin{equation}
    C^{-1} = \frac{N_m - N_{\rm data} - 2}{N_m -1} [C^{\rm tot}]^{-1} \ .
    \label{eq:hartlap}
\end{equation}
While this correction factor remains negligible for cases that consist of only the 2PCF, when combining it with the observed 3PCF it becomes non-trivial. For example, the data vector defined in Eq.~\ref{eq:data_vector} carries a Hartlap factor of 0.5586. While this method provides a better estimation of the covariance, it also highlights the need for increasingly large numbers of mock catalogs needed to model covariance for higher order statistics.

\section{Parameter Forecasts}
\label{sec:cosmo_forecasts}

    \begin{figure*}[!t]
        \centering
    	\includegraphics[width=\textwidth]{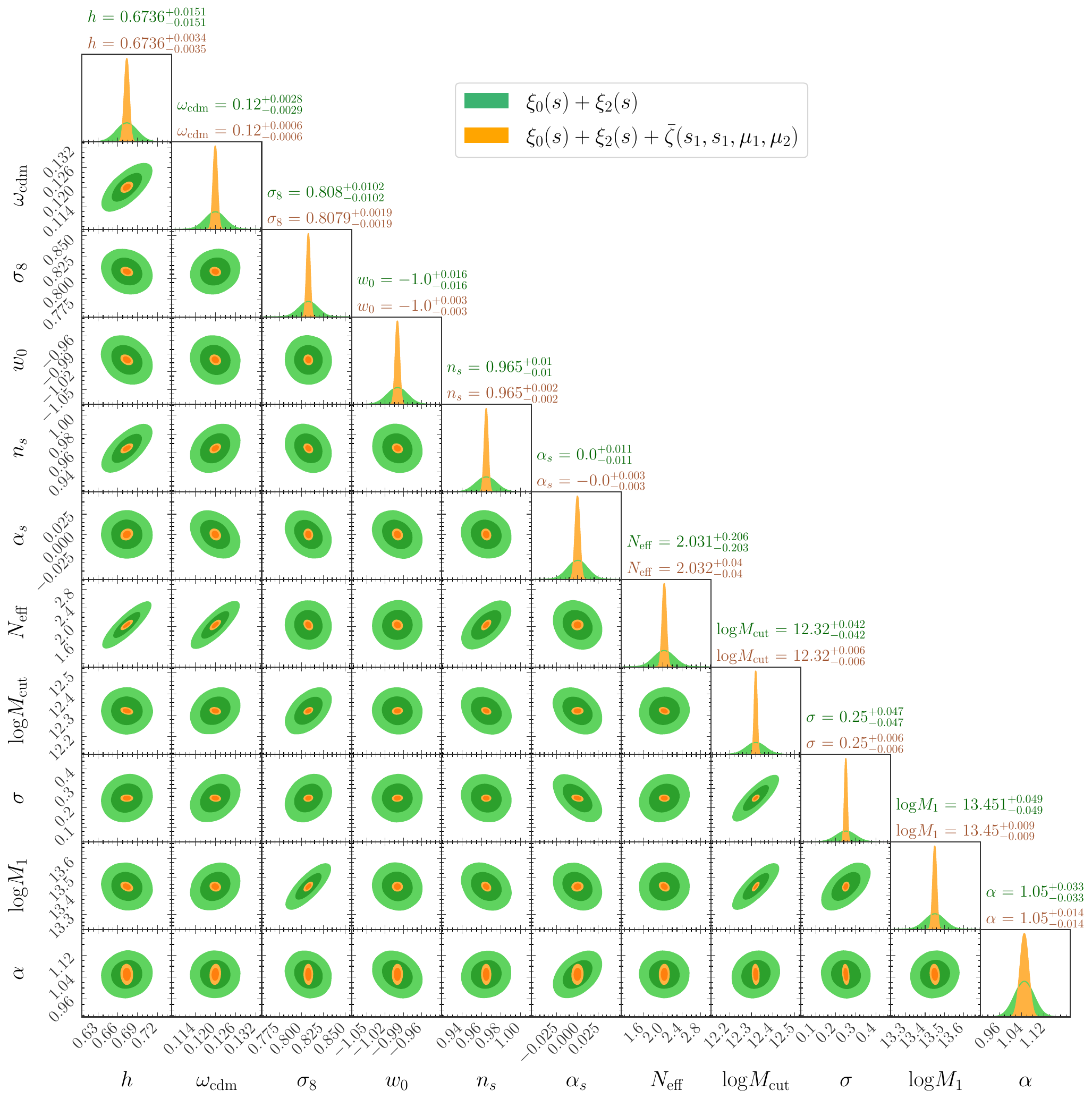}
    	\caption{Marginalized distributions of the parameters $\theta$ for the case where the data vector consists of only the 2PCF monopole and quadrupole (green) and where the data vector includes also the LOS decomposed 3PCF (orange) for Roman ELGs. The dark and light contours represent the 1$\sigma$ and 2$\sigma$ contours respectively. The most probable values and $1\sigma$ CL are labeled for each parameter above the respective panel in the appropriate color. This figure was generated using {\tt pygtc} \citep{bocquet2016pygtc}.}
        \label{fig:post_23_comp}
    \end{figure*}

To assess the constraining power of the data vector, we perform parameter inference using a Markov Chain Monte Carlo (MCMC) algorithm implemented with the \texttt{emcee} package \citep{foreman2013emcee}. The mock data vector is taken to be the fiducial model prediction, \(\widetilde{O}^{\rm fid}\). By construction, the linear model in Eq.~\ref{eq:model_linear} exactly reproduces this data vector at \(\theta=\theta^{\rm fid}\), so the best-fit residual vanishes.

For a fixed covariance and the linear response model defined above, the Gaussian likelihood is
\begin{equation}
    \label{eq:log_like}
    -2\log\mathcal{L}(\theta)
    =
    \left[\delta\theta\,J\right]^T
    C^{-1}
    \left[\delta\theta\,J\right],
\end{equation}
where \(J=P^{-1}\Delta\) is the response matrix defined in Section~\ref{sec:model_predictions}. The fiducial data vector cancels between the mock data and the model prediction. For a linear model with fixed covariance, this likelihood is equivalent to a Fisher forecast; we use MCMC as a convenient way to sample the posterior and visualize marginalized parameter constraints.

We adopt the effective covariance described in Section~\ref{sec:covariance}, corresponding to the volume of a full \AbacusSummit box. The resulting constraints should therefore be interpreted as fixed-volume forecasts rather than as predictions for a specific DESI or Roman survey sample. Under the assumption that the covariance scales approximately as \(1/V\) and that external priors are not dominant, the relative improvement obtained by adding the 3PCF is expected to be only weakly dependent on the chosen effective volume.

\begin{table*}[ht]
\centering
\scriptsize
\renewcommand{\arraystretch}{1.5}
\resizebox{\textwidth}{!}{%
\begin{tabular}{l c c c c c c c c c c c c c}
Data Vector Components &
$N_{\rm data}$ &
$({N_m - N_{\rm data} - 2})/({N_m - 1})$ &
$\sigma_{h}$ &
$\sigma_{\omega_{\mathrm{cdm}}}$ &
$\sigma_{\sigma_8}$ &
$\sigma_{w_0}$ &
$\sigma_{n_s}$ &
$\sigma_{\alpha_s}$ &
$\sigma_{N_{\mathrm{eff}}}$ &
$\sigma_{\mathrm{log} M_{\mathrm{cut}}}$ &
$\sigma_{\sigma}$ &
$\sigma_{\mathrm{log} M_1}$ &
$\sigma_{\alpha}$ \\
\hline
\hline
$\xi_0(s)$ &
16 &
0.983 &
0.0215 &
0.0055 &
0.0222 &
0.0276 &
0.0173 &
0.023 &
0.32 &
0.0849 &
0.0633 &
0.1159 &
0.2475 \\

$\xi_0(s) + \xi_2(s)$ &
32 &
0.967 &
0.0151 &
0.0029 &
0.0102 &
0.0159 &
0.0101 &
0.0114 &
0.2046 &
0.0422 &
0.047 &
0.049 &
0.0333 \\

$\bar{\zeta}(s_1,s_2,\mu_1,\mu_2)$ &
408 &
0.5906 &
0.0035 &
0.0006 &
0.0019 &
0.0032 &
0.0018 &
0.0029 &
0.0404 &
0.0081 &
0.0065 &
0.011 &
0.0265 \\

$\zeta_{0}(s_1,s_2)$ &
136 &
0.8629 &
0.0127 &
0.0021 &
0.0071 &
0.0115 &
0.0072 &
0.009 &
0.1498 &
0.024 &
0.0205 &
0.0376 &
0.1222 \\

$\xi_0(s) + \xi_2(s) + \bar{\zeta}(s_1,s_2,\mu_1,\mu_2)$ &
440 &
0.5586 &
0.0035 &
0.0006 &
0.0019 &
0.0032 &
0.0018 &
0.0026 &
0.0401 &
0.0061 &
0.0061 &
0.0092 &
0.0143 \\

$\xi_0(s) + \xi_2(s) + \zeta_{0}(s_1,s_2)$ &
168 &
0.8308 &
0.0096 &
0.0016 &
0.0054 &
0.009 &
0.0054 &
0.0065 &
0.1194 &
0.0154 &
0.0145 &
0.0241 &
0.025 \\

\hline
\hline

\end{tabular}
}

\caption{The number of data points, Hartlap factor, and 1$\sigma$ contraints on the model parameters for different combinations of the 2PCF and 3PCF corresponding to the Roman ELG sample.}
\label{tab:summary_tab_roman}
\end{table*}

\begin{table*}[ht]
\centering
\scriptsize
\renewcommand{\arraystretch}{1.5}
\resizebox{\textwidth}{!}{%
\begin{tabular}{l c c c c c c c c c c c c c}
Data Vector Components &
$N_{\rm data}$ &
$({N_m - N_{\rm data} - 2})/({N_m - 1})$ &
$\sigma_{h}$ &
$\sigma_{\omega_{\mathrm{cdm}}}$ &
$\sigma_{\sigma_8}$ &
$\sigma_{w_0}$ &
$\sigma_{n_s}$ &
$\sigma_{\alpha_s}$ &
$\sigma_{N_{\mathrm{eff}}}$ &
$\sigma_{\mathrm{log} M_{\mathrm{cut}}}$ &
$\sigma_{\sigma}$ &
$\sigma_{\mathrm{log} M_1}$ &
$\sigma_{\alpha}$ \\
\hline
\hline
$\xi_0(s)$ &
16 &
0.983 &
0.029 &
0.007 &
0.0115 &
0.0193 &
0.0156 &
0.0304 &
0.402 &
0.0507 &
0.0682 &
0.0905 &
0.2455 \\

$\xi_0(s) + \xi_2(s)$ &
32 &
0.967 &
0.0105 &
0.0022 &
0.0057 &
0.0127 &
0.0069 &
0.0097 &
0.1382 &
0.0226 &
0.0291 &
0.0222 &
0.0343 \\

$\bar{\zeta}(s_1,s_2,\mu_1,\mu_2)$ &
408 &
0.5906 &
0.0029 &
0.0006 &
0.0016 &
0.0027 &
0.0014 &
0.0024 &
0.0332 &
0.0066 &
0.0062 &
0.014 &
0.0331 \\

$\zeta_{0}(s_1,s_2)$ &
136 &
0.8629 &
0.0089 &
0.0021 &
0.005 &
0.0098 &
0.0046 &
0.008 &
0.1092 &
0.0194 &
0.0204 &
0.0449 &
0.1227 \\

$\xi_0(s) + \xi_2(s) + \bar{\zeta}(s_1,s_2,\mu_1,\mu_2)$ &
440 &
0.5586 &
0.0028 &
0.0006 &
0.0015 &
0.0026 &
0.0014 &
0.0023 &
0.033 &
0.0051 &
0.0061 &
0.0091 &
0.0166 \\

$\xi_0(s) + \xi_2(s) + \zeta_{0}(s_1,s_2)$ &
168 &
0.8308 &
0.0059 &
0.0013 &
0.0031 &
0.007 &
0.0036 &
0.0058 &
0.078 &
0.0115 &
0.0147 &
0.0155 &
0.0243\\

\hline
\hline

\end{tabular}
}

\caption{The number of data points, Hartlap factor, and 1$\sigma$ contraints on the model parameters for different combinations of the 2PCF and 3PCF corresponding to the DESI LRG sample.}
\label{tab:summary_tab_desi}
\end{table*}

The resulting marginalized posterior constraints are shown in Fig.~\ref{fig:post_23_comp} for Roman ELGs. The contours compare the constraints obtained from the 2PCF monopole and quadrupole alone with those obtained after adding the LOS-dependent 3PCF. Adding the 3PCF substantially tightens the posterior, with the largest gains appearing for the cosmological parameters. For most parameters, the marginalized uncertainties improve by factors of approximately \(2\)--\(4\), while the improvement for \(\sigma_8\) is larger. A quantitative summary of the marginalized errors and improvement factors is given in Table~\ref{tab:summary_tab_roman} for Roman ELGs, and in Table~\ref{tab:summary_tab_desi} for DESI LRGs.

\begin{figure*}[!t]
        \centering
    	\includegraphics[width=\textwidth]{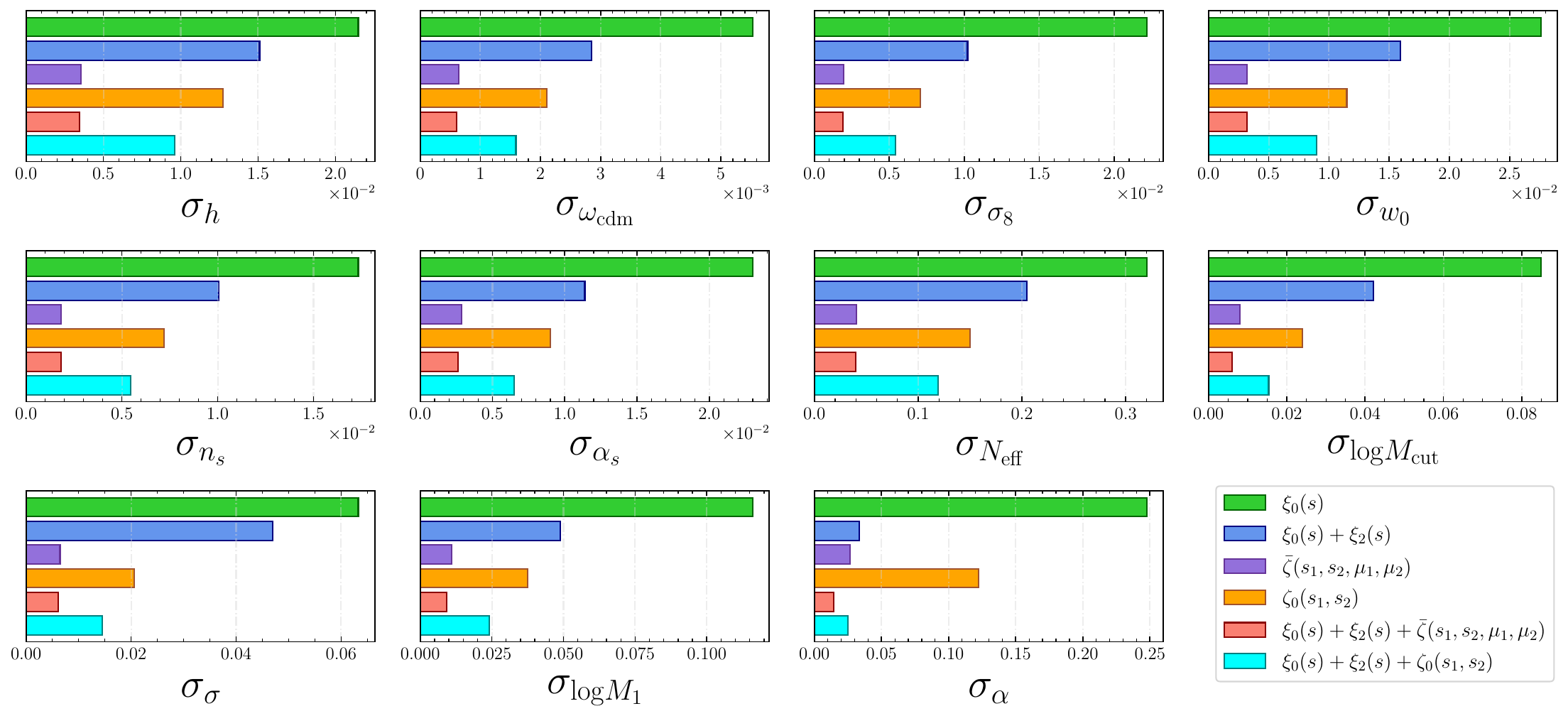}
    	\caption{The size of the 1$\sigma$ constraints on model parameters corresponding to the Roman ELG forecasts that include the 2PCF monopole (green), 2PCF monopole and quadrupole (blue), the $\mu$-averaged 3PCF (purple), the ``traditional'' 3PCF (orange), the combined 2PCF monopole, 2PCF quadrupole, and $\mu$-averaged 3PCF (red), and combined 2PCF monopole, 2PCF quadrupole, and ``traditional'' 3PCF (cyan).}
        \label{fig:summary_panel}
    \end{figure*}

To understand the contribution to the overall sensitivity from different elements of the data vector, we compare six different combinations of the 2PCF monopole, quadrupole, and both the traditional and LOS averaged 3PCF bases. Fig.~\ref{fig:summary_panel} shows the sensitivity to each parameter for all cases considered for Roman ELGs. These results reinforce the notion that the 3PCF adds considerable constraining power to the model. Moreover, it also shows the improvement of the LOS averaged 3PCF compared to the traditional 3PCF monopole. The exact numerical uncertainties corresponding to each case are given in Table~\ref{tab:summary_tab_roman} and Table~\ref{tab:summary_tab_desi}, showing complete results for both tracer samples.

\section{Validation of the Linear Model}
\label{sec:robustness}

    \begin{figure*}[!t]
        \centering
    	\includegraphics[width=\textwidth]{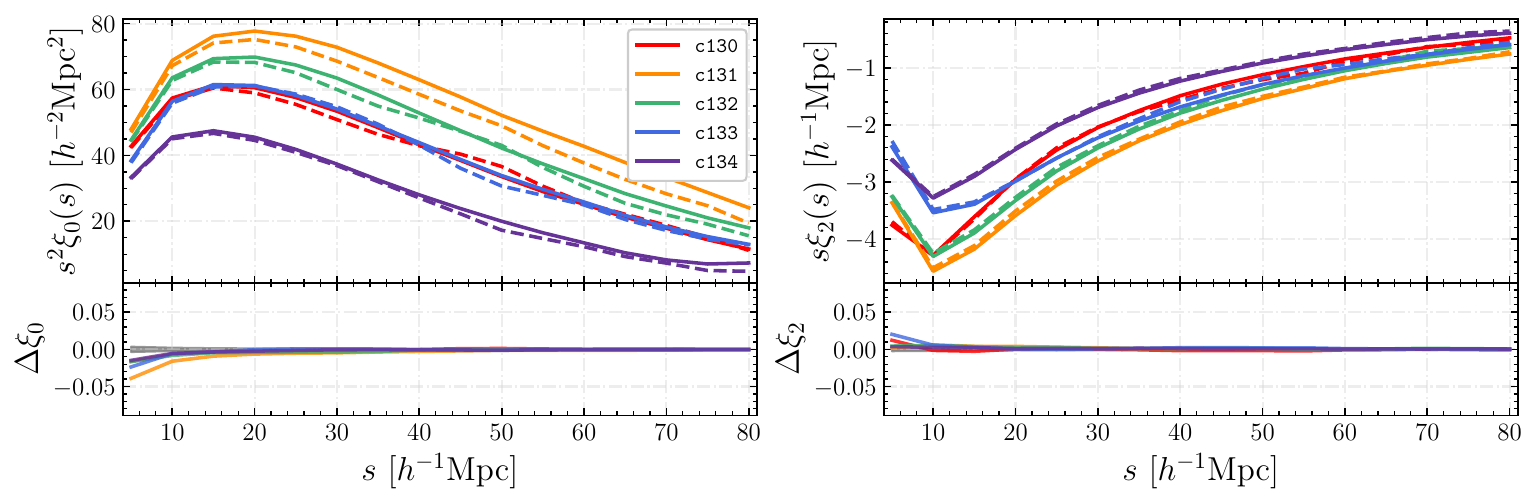}
    	\caption{Upper panel: The 2PCF, $\xi_{\ell}(s)$, of 5 cosmologies in the {\tt AbacusSummit} emulator grid at Roman ELG densities. Solid color lines represent the measured 2PCF, while dashed lines indicate the 2PCF as predicted by our Fisher model. Lower panel: The difference between the predicted versus measured 2PCF corresponding to each of the same 5 emulator grid cosmologies. The grey shaded band represents the 1$\sigma$ errors on the 2PCF in our fits.}
        \label{fig:xi_em_grid}
    \end{figure*}
    
    \begin{figure*}[!t]
        \centering
    	\includegraphics[width=\textwidth]{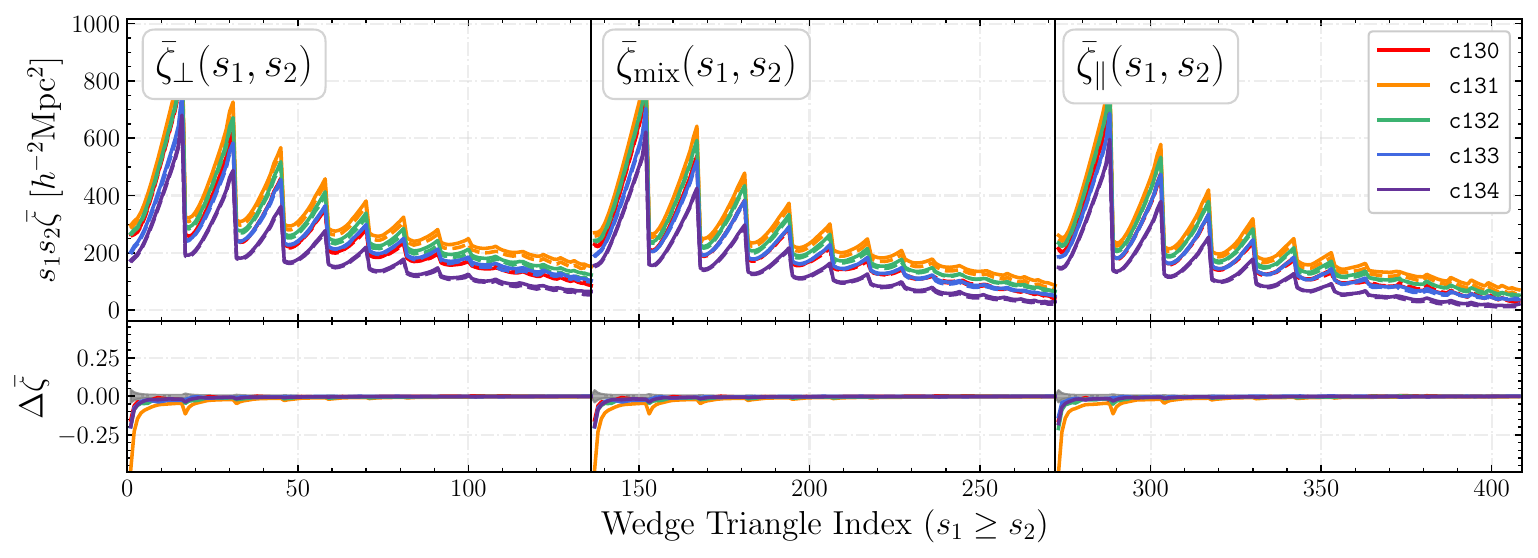}
    	\caption{Upper panel: The 3PCF, $\bar{\zeta}(s_1,s_2)$,  of 5 cosmologies in the {\tt AbacusSummit} emulator grid at Roman ELG densities. Solid color lines represent the measured 3PCF, while dashed lines indicate the 3PCF as predicted by our Fisher model. Lower panel: The difference between the predicted versus measured 3PCF corresponding to each of the same 5 emulator grid cosmologies. The grey shaded band represents the 1$\sigma$ errors on the 3PCF in our fits. Both panels are separated into the three unique $\mu$-configurations.}
        \label{fig:zeta_em_grid}
    \end{figure*}

The forecasts presented above rely on a linear expansion of the data vector around the fiducial cosmology. This approximation is expected to be most accurate close to the fiducial point, but it is important to test how well it reproduces clustering measurements away from the derivative grid used to construct the model.

To perform this test, we use five additional cosmologies from the \AbacusSummit emulator grid, labeled {\tt c130}--{\tt c134}. These cosmologies were chosen because they are the only available emulator-grid models that satisfy two requirements relevant for this validation. First, their parameter displacements lie within approximately three steps of the derivative simulations used to build the linear model. Second, they vary only parameters included in our 11-dimensional forecast parameter vector. Other available emulator-grid cosmologies vary additional parameters, such as \(w_a\) or \(\omega_b\), which are not included in the model used in this work.

For each of the five validation cosmologies, we measure the 2PCF and 3PCF using the same pipeline applied to the fiducial and offset simulations. We then compare these measurements to the predictions of the linear model in Eq.~\eqref{eq:model_linear}. The results are shown in Figs.~\ref{fig:xi_em_grid} and~\ref{fig:zeta_em_grid} for Roman ELGs. Solid curves show the direct measurements from the simulations, while dashed curves show the corresponding linear-model predictions. The lower panels show the residuals between the prediction and the measured clustering statistic, with the grey band indicating the \(1\sigma\) statistical uncertainty for a single full \AbacusSummit box.

The agreement is reasonable on intermediate and large scales, but the comparison also shows clear limitations of the linear approximation. On the smallest scales included in the analysis, the residuals can exceed the statistical uncertainty of one \AbacusSummit box. Moreover, the discrepancies are not purely random: the linear model tends to underpredict the measured monopole and 3PCF amplitudes, while it tends to overpredict the quadrupole. This behavior indicates that the parameter dependence of the clustering statistics is not perfectly captured by a first-order expansion over the full range spanned by these validation cosmologies.

For a production analysis of real survey data, such modeling residuals would need to be addressed with a more accurate emulator, a more local derivative scheme, additional simulations, or more conservative scale cuts. In this work, however, our goal is more limited: we use the linear model to estimate the relative improvement obtained by adding the LOS-dependent 3PCF to the 2PCF. The posterior constraints from the combined 2PCF+3PCF data vector are substantially tighter than the full range tested by the validation cosmologies. Therefore, the region of parameter space most relevant for the joint 2PCF+3PCF forecasts lies much closer to the fiducial model than the extreme validation points shown here. Within this smaller region, the linear approximation is expected to be adequate for the forecast-level comparison performed in this paper.

We therefore interpret this validation as supporting the use of the linear model for the present forecasts, while also highlighting that the same approach would not by itself be sufficient for a precision analysis of real survey data over a broad parameter volume.

\section{Scale and configuration dependence of the results}
\label{sec:scale_dep}

    \begin{figure}[!t]
        \centering
    	\includegraphics[width=0.45\textwidth]{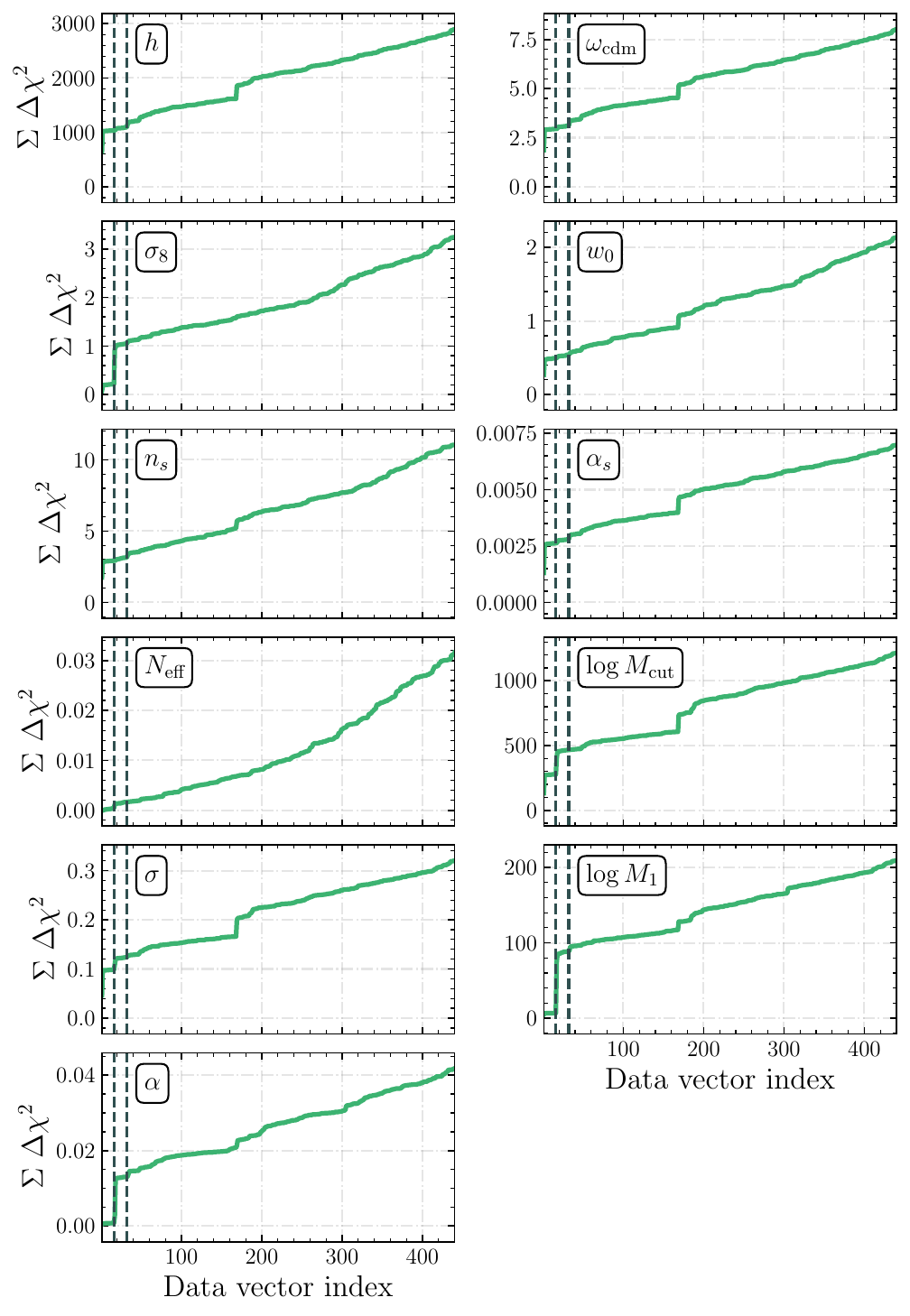}
    	\caption{The cumulative statistical difference, $\Sigma \chi^2$, between the fiducial case and a case where the parameter has been varied by 1$\%$ for Roman ELGs. Each panel corresponds to on of the model parameters. The dashed vertical lines indicate the boundaries between the 2PCF monopole, 2PCF quadrupole, and 3PCF.}
        \label{fig:cum_chi2}
    \end{figure}

To understand which parts of the data vector contribute to these improvements, we compute a cumulative \(\chi^2\)-like diagnostic as a function of the ordered data-vector index. For each parameter, we evaluate the model difference produced by a \(1\%\) displacement from the fiducial value,
\begin{equation}
    \delta O = \widetilde{O}(\theta)-\widetilde{O}^{\rm fid}.
\end{equation}
For each value of \(k\), we restrict both the model-difference vector and the covariance matrix to the first \(k\) elements of the ordered data vector. We then compute
\begin{equation}
    \label{eq:cum_chi2}
    \chi^2(k)
    =
    \delta O_{(k)}^{T}
    C_{(k)}^{-1}
    \delta O_{(k)} ,
\end{equation}
where \(\delta O_{(k)}\) is the truncated data-vector difference and \(C_{(k)}\) is the corresponding \(k\times k\) covariance submatrix. Thus, for each \(k\), the covariance is first restricted to the same subset of bins and then inverted.

Because this diagnostic depends on the chosen ordering of the data vector, it should not be interpreted as a unique localization of information. It is intended only to indicate whether the constraining power accumulates gradually across many bins or is dominated by a small number of configurations.

The results are shown in Fig.~\ref{fig:cum_chi2} for Roman ELGs. The cumulative curves indicate that the information added by the 3PCF is distributed broadly across triangle configurations rather than being localized to a single scale or configuration. Although the 3PCF bins are highly correlated, the large number of distinct configurations leads to a substantial cumulative increase in constraining power.

For parameters such as \(\sigma_8\), \(\sigma\), \(\alpha\), \(M_{\mathrm{cut}}\), and \(M_1\), the addition of quadrupole information yields a significant improvement over the monopole-only case. Additional gains are observed when including the 3PCF, particularly for parameters such as \(h\), \(\sigma\), \(\omega_{\mathrm{cdm}}\), and \(M_{\mathrm{cut}}\), where configurations sensitive to line-of-sight anisotropy provide complementary information.

\begin{figure*}[!t]
        \centering
    	\includegraphics[width=\textwidth]{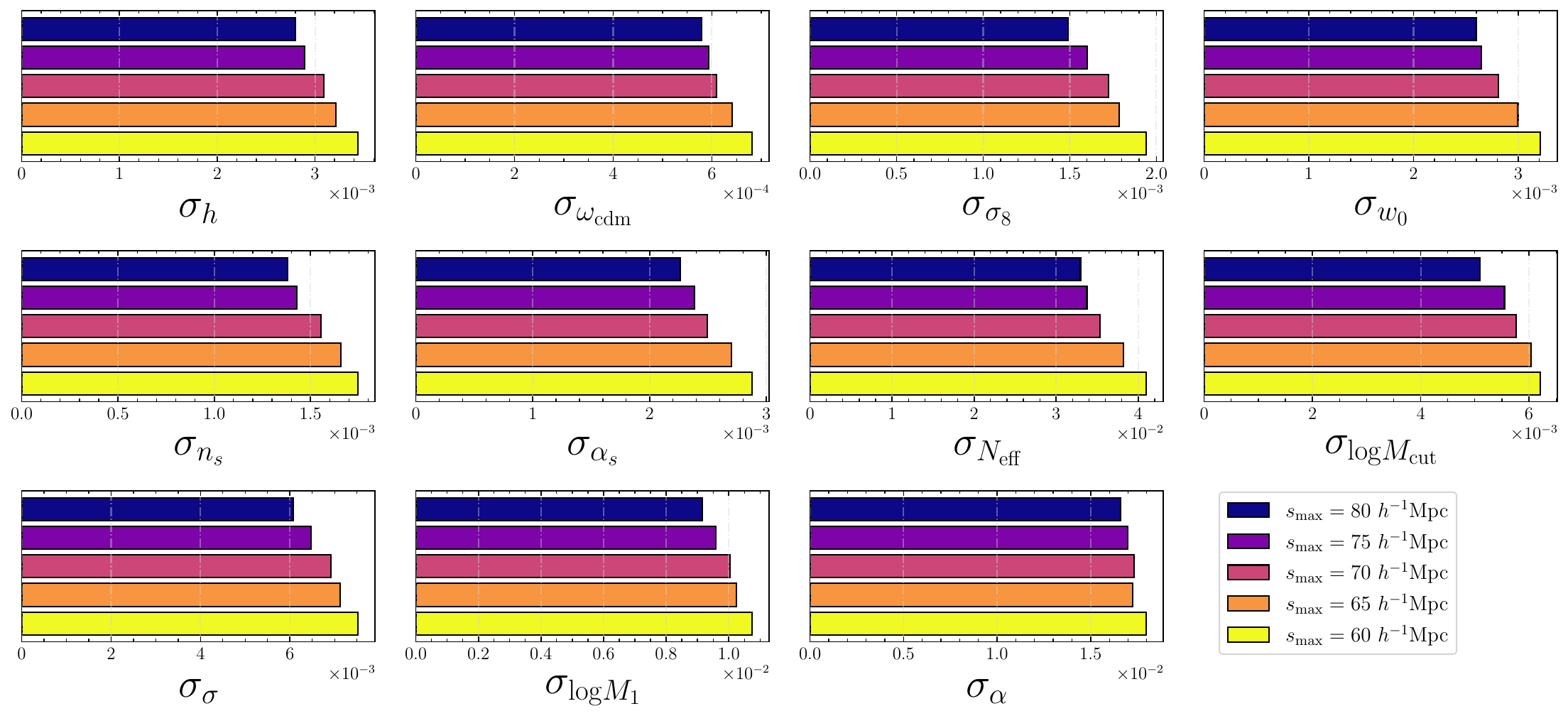}
    	\caption{The size of the Roman ELG 1$\sigma$ constraints on model parameters corresponding to forecasts that vary the maximum scale, $s_{\rm max}$ from 60 $h^{-1}$Mpc to the fiducial value of 80 $h^{-1}$Mpc. All cases correspond to a data vector with components indicated by Eq.~\ref{eq:data_vector} and a minimum scale of $s_{\rm min}$ = 5 $h^{-1}$Mpc.}
        \label{fig:summary_panel_smax}
    \end{figure*}

\begin{figure*}[!t]
    \centering
    \includegraphics[width=\textwidth]{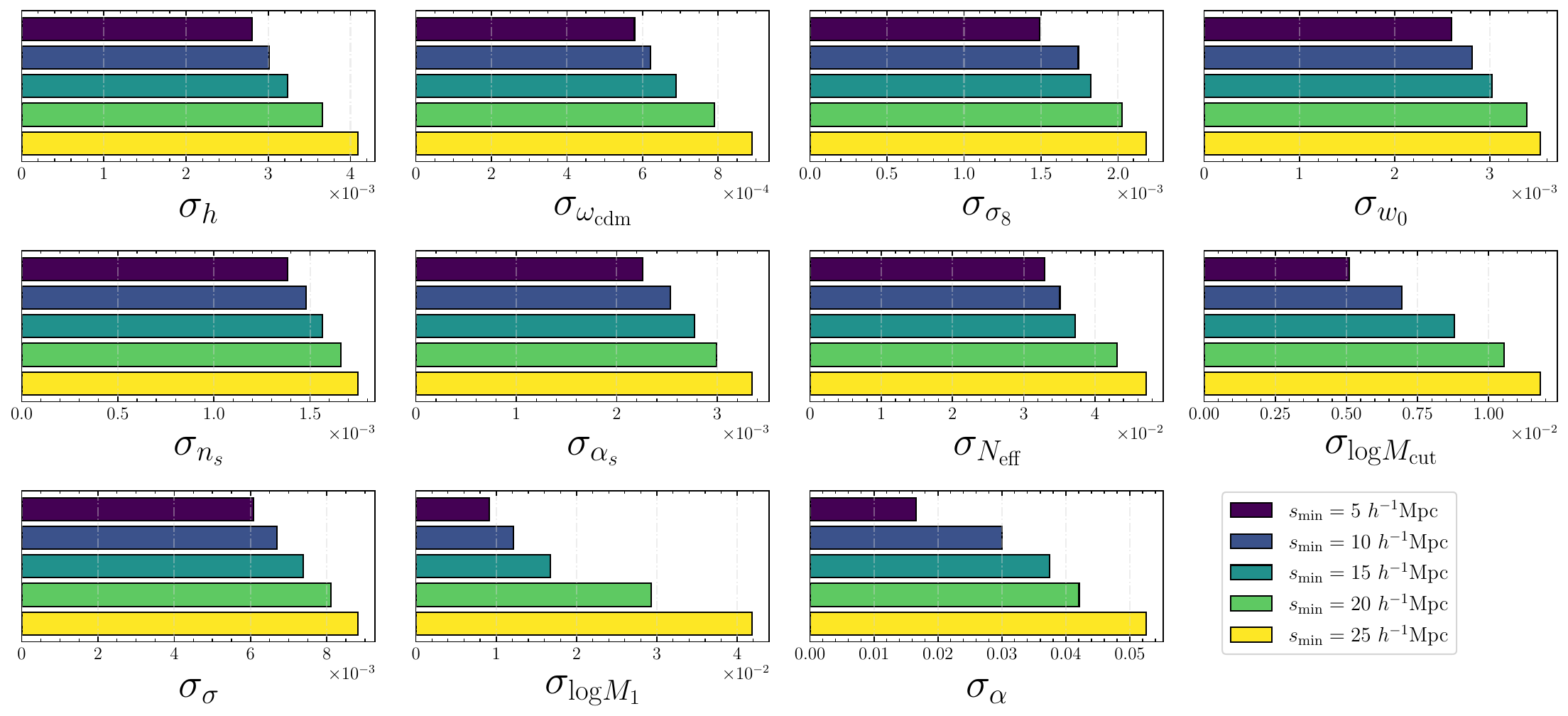}
    \caption{The size of the Roman ELG 1$\sigma$ constraints on model parameters corresponding to forecasts that vary the minimum scale, $s_{\rm min}$ from the fiducial value of 5 $h^{-1}$Mpc to 25 $h^{-1}$Mpc. All cases correspond to a data vector with components indicated by Eq.~\ref{eq:data_vector} and a maximum scale of $s_{\rm max}$ = 80 $h^{-1}$Mpc.}
    \label{fig:summary_panel_smin}
\end{figure*}

We can explicitly test the dependence of the model on both large and small scales. Fig.~\ref{fig:summary_panel_smax} and Fig.~\ref{fig:summary_panel_smin} show the effects of varying the maximum and minimum scales respectively for the Roman ELG sample. The improvement in our measurements with increasing $s_{\rm max}$ is nearly linear, again showing how our sensitivity comes not from an individual feature, but in a cumulative increase in clustering information. Conversely, varying $s_{\rm min}$ shows how, especially for the HOD parameters, the smallest scales are increasingly sensitive. This is promising, as it allows us to probe more statistically relevant information by pushing our observations to smaller scales. However, we remain wary of including clustering properties at scales smaller than these as they are highly sensitive to systematics in galaxy redshift surveys.

\section{Discussion and Conclusions}
\label{sec:discussion_conclusions}

We have investigated the cosmological information content of small-scale two- and three-point galaxy clustering in configuration space using Roman-like ELG and DESI-like LRG mock catalogs constructed from the \AbacusSummit simulations. The main novelty of this work is the combination of a compact line-of-sight-dependent configuration-space 3PCF statistic, fast triangle counting with \textsc{TriCo}, and simulation-based forecasts in which uncertainty in the galaxy--halo connection is explicitly marginalized over. Our main result is that this compressed LOS-dependent 3PCF measurement provides substantial additional constraining power beyond the 2PCF monopole and quadrupole. For the Roman-like ELG sample, adding the 3PCF tightens the marginalized constraint on $\sigma_8$ by a factor of approximately five relative to the 2PCF-only case. As shown in Tables 2 and 3, broadly similar improvements are obtained for the DESI-like LRG sample, indicating that the gain is not unique to the high-density Roman-like ELG case.

The improvement from the 3PCF is not driven by a single scale or by a small number of isolated triangle configurations. Instead, the gain accumulates across many configurations of the LOS-dependent 3PCF data vector. This behavior is encouraging because it suggests that the additional constraining power is a genuine higher-order clustering effect rather than a response to a localized feature in the data vector. The LOS dependence is also important: retaining information about the orientation of triangle configurations relative to the line-of-sight captures anisotropic clustering information that is only partially accessible in LOS-averaged measurements. In particular, the full LOS-dependent statistic outperforms the LOS-averaged 3PCF, showing that the orientation dependence of triangle configurations carries useful cosmological information.

A key feature of this analysis is that the HOD parameters are treated as nuisance parameters and marginalized over rather than fixed. This is essential for interpreting small-scale clustering constraints, where changes in the galaxy--halo connection can mimic or obscure cosmological parameter dependence. The fact that the 3PCF continues to improve cosmological constraints after HOD marginalization indicates that higher-order clustering helps break degeneracies between cosmology and galaxy bias. In this sense, the 3PCF is useful not only because it adds more data points, but because it probes aspects of the galaxy distribution that respond differently to cosmological and HOD parameters.

The forecasts presented here should not be interpreted as predictions for the exact constraining power of either the Roman ELG sample or the DESI Y1 LRG sample. Our covariance corresponds to an effective periodic-box volume rather than to a realistic survey geometry, and it does not include the full set of observational effects present in survey data, such as the angular mask, radial selection, imaging systematics, fiber assignment, or redshift failures. Instead, our goal has been to quantify the relative information gain from adding a LOS-dependent 3PCF measurement at fixed effective volume. Under the assumption that the dominant covariance terms scale approximately with inverse volume, these relative improvements should provide a useful guide to the value of including 3PCF information in future full-shape analyses.

The validation tests using additional \AbacusSummit emulator-grid cosmologies show that the linear response model captures the broad parameter dependence of the 2PCF and 3PCF, but they also reveal limitations. In particular, the model residuals can exceed the statistical uncertainty of a single full \AbacusSummit box on the smallest scales, and the residuals show coherent trends for some statistics. This level of modeling error would need to be addressed in a precision analysis of real survey data, for example through a more complete emulator, additional simulations, or more conservative scale cuts. For the forecast-level comparison performed here, the approximation is most reliable within the smaller parameter region selected by the joint 2PCF+3PCF constraints. The validation tests therefore caution against interpreting the small-scale 2PCF-only Fisher forecasts as fully robust absolute constraints, while still supporting the relative conclusion that the LOS-dependent 3PCF provides substantial additional information.

Beyond the specific forecasts shown here, these results support the broader use of higher-order configuration-space statistics in full-shape cosmological analyses. Realistic survey analyses must marginalize over both galaxy--halo uncertainties and observational nuisance parameters. Higher-order statistics can help constrain these nuisance degrees of freedom because they respond differently to cosmology, galaxy bias, and observational systematics. This makes the 3PCF a promising complement to standard 2PCF-based full-shape measurements, especially on mildly and fully non-linear scales where much of the available clustering information is otherwise difficult to use.

In summary, we find that small-scale LOS-dependent 3PCF measurements can significantly increase the cosmological sensitivity of configuration-space galaxy clustering analyses. For Roman-like ELGs, adding the LOS-dependent 3PCF to the 2PCF monopole and quadrupole improves the marginalized $\sigma_8$ constraint by a factor of approximately five, with substantial gains also seen for other cosmological and HOD parameters. The DESI-like LRG forecasts show the same qualitative behavior, demonstrating that the usefulness of the statistic extends beyond a single tracer sample. This work provides a step toward simulation-based full-shape analyses that combine two- and three-point information while marginalizing over the galaxy--halo connection. A natural next step is to extend the same framework to realistic survey geometries and to additional samples, where the balance between number density, redshift, non-linear growth, and observational systematics may change the relative value of the 3PCF.

\section*{Acknowledgments}

Z. Brown and L. Samushia gratefully acknowledge funding from NASA Grant $\#$80NSSC24M0021. L Samushia also gratefully acknowledges support from the NASA ROSES grant 12-EUCLID12-0004.

The authors thank N. Magnelli for the production of the \AbacusSummit mock catalogs and G. Khomeriki, D. Modebadze, H. Randall, and R. Demina for helpful discussions.

\bibliographystyle{aasjournal}

\bibliography{bib}

@article{cole20052df,
  title={The 2dF Galaxy Redshift Survey: power-spectrum analysis of the final data set and cosmological implications},
  author={Cole, Shaun and Percival, Will J and Peacock, John A and Norberg, Peder and Baugh, Carlton M and Frenk, Carlos S and Baldry, Ivan and Bland-Hawthorn, Joss and Bridges, Terry and Cannon, Russell and others},
  journal={Monthly Notices of the Royal Astronomical Society},
  volume={362},
  number={2},
  pages={505--534},
  year={2005},
  publisher={The Royal Astronomical Society}
}

@article{parkinson2012wigglez,
  title={The WiggleZ dark energy survey: final data release and cosmological results},
  author={Parkinson, David and Riemer-S{\o}rensen, Signe and Blake, Chris and Poole, Gregory B and Davis, Tamara M and Brough, Sarah and Colless, Matthew and Contreras, Carlos and Couch, Warrick and Croom, Scott and others},
  journal={Physical Review D—Particles, Fields, Gravitation, and Cosmology},
  volume={86},
  number={10},
  pages={103518},
  year={2012},
  publisher={APS}
}

@article{alam2017clustering,
  title={The clustering of galaxies in the completed SDSS-III Baryon Oscillation Spectroscopic Survey: cosmological analysis of the DR12 galaxy sample},
  author={Alam, Shadab and Ata, Metin and Bailey, Stephen and Beutler, Florian and Bizyaev, Dmitry and Blazek, Jonathan A and Bolton, Adam S and Brownstein, Joel R and Burden, Angela and Chuang, Chia-Hsun and others},
  journal={Monthly Notices of the Royal Astronomical Society},
  volume={470},
  number={3},
  pages={2617--2652},
  year={2017},
  publisher={Oxford University Press}
}

@article{adame2025desi7,
  title={DESI 2024 VII: cosmological constraints from the full-shape modeling of clustering measurements},
  author={Adame, AG and Aguilar, J and Ahlen, S and Alam, S and Alexander, DM and Prieto, C Allende and Alvarez, M and Alves, O and Anand, A and Andrade, U and others},
  journal={Journal of Cosmology and Astroparticle Physics},
  volume={2025},
  number={07},
  pages={028},
  year={2025},
  publisher={IOP Publishing}
}

@article{guth1982fluctuations,
  title={Fluctuations in the new inflationary universe},
  author={Guth, Alan H and Pi, So-Young},
  journal={Physical Review Letters},
  volume={49},
  number={15},
  pages={1110},
  year={1982},
  publisher={APS}
}

@article{bardeen1983spontaneous,
  title={Spontaneous creation of almost scale-free density perturbations in an inflationary universe},
  author={Bardeen, James M and Steinhardt, Paul J and Turner, Michael S},
  journal={Physical Review D},
  volume={28},
  number={4},
  pages={679},
  year={1983},
  publisher={APS}
}

@article{keitel2011constrained,
  title={Constrained probability distributions of correlation functions},
  author={Keitel, David and Schneider, Peter},
  journal={Astronomy \& Astrophysics},
  volume={534},
  pages={A76},
  year={2011},
  publisher={EDP Sciences}
}

@article{slepian2015computing,
  title={Computing the three-point correlation function of galaxies in $$\backslash$mathcal $\{$O$\}$(N\^{} 2) $ time},
  author={Slepian, Zachary and Eisenstein, Daniel J},
  journal={Monthly Notices of the Royal Astronomical Society},
  volume={454},
  number={4},
  pages={4142--4158},
  year={2015},
  publisher={OUP}
}

@article{blake2011wigglez,
  title={The WiggleZ Dark Energy Survey: testing the cosmological model with baryon acoustic oscillations at z= 0.6},
  author={Blake, Chris and Davis, Tamara and Poole, Gregory B and Parkinson, David and Brough, Sarah and Colless, Matthew and Contreras, Carlos and Couch, Warrick and Croom, Scott and Drinkwater, Michael J and others},
  journal={Monthly Notices of the Royal Astronomical Society},
  volume={415},
  number={3},
  pages={2892--2909},
  year={2011},
  publisher={Blackwell Publishing Ltd Oxford, UK}
}

@article{bautista2021completed,
  title={The completed SDSS-IV extended Baryon Oscillation Spectroscopic Survey: measurement of the BAO and growth rate of structure of the luminous red galaxy sample from the anisotropic correlation function between redshifts 0.6 and 1},
  author={Bautista, Julian E and Paviot, Romain and Vargas Magana, Mariana and de La Torre, Sylvain and Fromenteau, Sebastien and Gil-Marín, Hector and Ross, Ashley J and Burtin, Etienne and Dawson, Kyle S and Hou, Jiamin and others},
  journal={Monthly Notices of the Royal Astronomical Society},
  volume={500},
  number={1},
  pages={736--762},
  year={2021},
  publisher={Oxford University Press}
}

@article{dumerchat2022baryon,
  title={Baryon acoustic oscillations from a joint analysis of the large-scale clustering in Fourier and configuration space},
  author={Dumerchat, Tyann and Bautista, Julian E},
  journal={Astronomy \& Astrophysics},
  volume={667},
  pages={A80},
  year={2022},
  publisher={EDP Sciences}
}

@article{montesano2010new,
  title={A new model for the full shape of the large-scale power spectrum},
  author={Montesano, Francesco and Sánchez, Ariel G and Phleps, Stefanie},
  journal={Monthly Notices of the Royal Astronomical Society},
  volume={408},
  number={4},
  pages={2397--2412},
  year={2010},
  publisher={Blackwell Publishing Ltd Oxford, UK}
}

@article{philcox2022boss,
  title={BOSS DR12 full-shape cosmology: $\Lambda$ CDM constraints from the large-scale galaxy power spectrum and bispectrum monopole},
  author={Philcox, Oliver HE and Ivanov, Mikhail M},
  journal={Physical Review D},
  volume={105},
  number={4},
  pages={043517},
  year={2022},
  publisher={APS}
}

@article{adame2025desi5,
  title={DESI 2024 V: Full-shape galaxy clustering from galaxies and quasars},
  author={Adame, AG and Aguilar, J and Ahlen, S and Alam, S and Alexander, DM and Alvarez, M and Alves, O and Anand, A and Andrade, U and Armengaud, E and others},
  journal={Journal of Cosmology and Astroparticle Physics},
  volume={2025},
  number={09},
  pages={008},
  year={2025},
  publisher={IOP Publishing}
}

@article{maksimova2021abacussummit,
  title={AbacusSummit: a massive set of high-accuracy, high-resolution N-body simulations},
  author={Maksimova, Nina A and Garrison, Lehman H and Eisenstein, Daniel J and Hadzhiyska, Boryana and Bose, Sownak and Satterthwaite, Thomas P},
  journal={Monthly Notices of the Royal Astronomical Society},
  volume={508},
  number={3},
  pages={4017--4037},
  year={2021},
  publisher={Oxford University Press}
}

@article{garrison2021abacus,
  title={The abacus cosmological N-body code},
  author={Garrison, Lehman H and Eisenstein, Daniel J and Ferrer, Douglas and Maksimova, Nina A and Pinto, Philip A},
  journal={Monthly Notices of the Royal Astronomical Society},
  volume={508},
  number={1},
  pages={575--596},
  year={2021},
  publisher={Oxford University Press}
}

@article{aghamousa2016desi1,
  title={The DESI experiment part I: science, targeting, and survey design},
  author={Aghamousa, Amir and Aguilar, Jessica and Ahlen, Steve and Alam, Shadab and Allen, Lori E and Prieto, Carlos Allende and Annis, James and Bailey, Stephen and Balland, Christophe and Ballester, Otger and others},
  journal={arXiv preprint arXiv:1611.00036},
  year={2016}
}

@article{aghamousa2016desi2,
  title={The DESI experiment part II: instrument design},
  author={Aghamousa, Amir and Aguilar, Jessica and Ahlen, Steve and Alam, Shadab and Allen, Lori E and Prieto, Carlos Allende and Annis, James and Bailey, Stephen and Balland, Christophe and Ballester, Otger and others},
  journal={arXiv preprint arXiv:1611.00037},
  year={2016}
}

@article{levi2013desi,
  title={The DESI Experiment, a whitepaper for Snowmass 2013},
  author={Levi, Michael and Bebek, Chris and Beers, Timothy and Blum, Robert and Cahn, Robert and Eisenstein, Daniel and Flaugher, Brenna and Honscheid, Klaus and Kron, Richard and Lahav, Ofer and others},
  journal={arXiv preprint arXiv:1308.0847},
  year={2013}
}

@article{prieto2020preliminary,
  title={Preliminary Target Selection for the DESI Milky Way Survey (MWS)},
  author={Prieto, Carlos Allende and Cooper, Andrew P and Dey, Arjun and G{\"a}nsicke, Boris T and Koposov, Sergey E and Li, Ting and Manser, Christopher and Nidever, David L and Rockosi, Constance and Wang, Mei-Yu and others},
  journal={Research Notes of the AAS},
  volume={4},
  number={10},
  pages={188},
  year={2020},
  publisher={The American Astronomical Society}
}

@article{ruiz2020preliminary,
  title={Preliminary target selection for the DESI bright galaxy survey (BGS)},
  author={Ruiz-Macias, Omar and Zarrouk, Pauline and Cole, Shaun and Norberg, Peder and Baugh, Carlton and Brooks, David and Dey, Arjun and Duan, Yutong and Eftekharzadeh, Sarah and Eisenstein, Daniel J and others},
  journal={Research Notes of the AAS},
  volume={4},
  number={10},
  pages={187},
  year={2020},
  publisher={The American Astronomical Society}
}

@article{zhou2020preliminary,
  title={Preliminary target selection for the DESI luminous red galaxy (LRG) sample},
  author={Zhou, Rongpu and Newman, Jeffrey A and Dawson, Kyle S and Eisenstein, Daniel J and Brooks, David D and Dey, Arjun and Dey, Biprateep and Duan, Yutong and Eftekharzadeh, Sarah and Gazta{\~n}aga, Enrique and others},
  journal={Research Notes of the AAS},
  volume={4},
  number={10},
  pages={181},
  year={2020},
  publisher={The American Astronomical Society}
}

@article{zhou2023target,
  title={Target selection and validation of DESI luminous red galaxies},
  author={Zhou, Rongpu and Dey, Biprateep and Newman, Jeffrey A and Eisenstein, Daniel J and Dawson, K and Bailey, S and Berti, A and Guy, J and Lan, Ting-Wen and Zou, H and others},
  journal={The Astronomical Journal},
  volume={165},
  number={2},
  pages={58},
  year={2023},
  publisher={IOP Publishing}
}

@article{yeche2020preliminary,
  title={Preliminary Target Selection for the DESI Quasar (QSO) Sample},
  author={Y{\`e}che, Christophe and Palanque-Delabrouille, Nathalie and Claveau, Charles-Antoine and Brooks, David D and Chaussidon, Edmond and Davis, Tamara M and Dawson, Kyle S and Dey, Arjun and Duan, Yutong and Eftekharzadeh, Sarah and others},
  journal={Research Notes of the AAS},
  volume={4},
  number={10},
  pages={179},
  year={2020},
  publisher={The American Astronomical Society}
}

@article{lan2023desi,
  title={The DESI Survey Validation: Results from Visual Inspection of Bright Galaxies, Luminous Red Galaxies, and Emission-line Galaxies},
  author={Lan, Ting-Wen and Tojeiro, R and Armengaud, E and Prochaska, J Xavier and Davis, TM and Alexander, David M and Raichoor, Anand and Zhou, Rongpu and Yeche, Christophe and Balland, C and others},
  journal={The Astrophysical Journal},
  volume={943},
  number={1},
  pages={68},
  year={2023},
  publisher={IOP Publishing}
}

@article{alexander2023desi,
  title={The DESI Survey Validation: Results from Visual Inspection of the Quasar Survey Spectra},
  author={Alexander, David M and Davis, Tamara M and Chaussidon, E and Fawcett, VA and Gonzalez-Morales, Alma X and Lan, Ting-Wen and Yeche, Christophe and Ahlen, S and Aguilar, JN and Armengaud, E and others},
  journal={The Astronomical Journal},
  volume={165},
  number={3},
  pages={124},
  year={2023},
  publisher={IOP Publishing}
}

@article{cooper2023overview,
  title={Overview of the DESI Milky Way survey},
  author={Cooper, Andrew P and Koposov, Sergey E and Prieto, Carlos Allende and Manser, Christopher J and Kizhuprakkat, Namitha and Myers, Adam D and Dey, Arjun and G{\"a}nsicke, Boris T and Li, Ting S and Rockosi, Constance and others},
  journal={The Astrophysical Journal},
  volume={947},
  number={1},
  pages={37},
  year={2023},
  publisher={IOP Publishing}
}

@article{hahn2023desi,
  title={The DESI bright galaxy survey: Final target selection, design, and validation},
  author={Hahn, ChangHoon and Wilson, Michael J and Ruiz-Macias, Omar and Cole, Shaun and Weinberg, David H and Moustakas, John and Kremin, Anthony and Tinker, Jeremy L and Smith, Alex and Wechsler, Risa H and others},
  journal={The Astronomical Journal},
  volume={165},
  number={6},
  pages={253},
  year={2023},
  publisher={IOP Publishing}
}

@article{raichoor2023target,
  title={Target selection and validation of DESI emission line galaxies},
  author={Raichoor, Anand and Moustakas, J and Newman, Jeffrey A and Karim, T and Ahlen, S and Alam, Shadab and Bailey, S and Brooks, D and Dawson, K and de la Macorra, A and others},
  journal={The Astronomical Journal},
  volume={165},
  number={3},
  pages={126},
  year={2023},
  publisher={IOP Publishing}
}

@article{chaussidon2023target,
  title={Target selection and validation of DESI quasars},
  author={Chaussidon, Edmond and Yeche, Christophe and Palanque-Delabrouille, Nathalie and Alexander, David M and Yang, Jinyi and Ahlen, Steven and Bailey, Stephen and Brooks, David and Cai, Zheng and Chabanier, Solene and others},
  journal={The Astrophysical Journal},
  volume={944},
  number={1},
  pages={107},
  year={2023},
  publisher={IOP Publishing}
}

@article{yuan2022abacushod,
  title={AbacusHOD: a highly efficient extended multitracer HOD framework and its application to BOSS and eBOSS data},
  author={Yuan, Sihan and Garrison, Lehman H and Hadzhiyska, Boryana and Bose, Sownak and Eisenstein, Daniel J},
  journal={Monthly Notices of the Royal Astronomical Society},
  volume={510},
  number={3},
  pages={3301--3320},
  year={2022},
  publisher={Oxford University Press}
}

@article{sinha2020corrfunc,
  title={corrfunc--a suite of blazing fast correlation functions on the CPU},
  author={Sinha, Manodeep and Garrison, Lehman H},
  journal={Monthly Notices of the Royal Astronomical Society},
  volume={491},
  number={2},
  pages={3022--3041},
  year={2020},
  publisher={Oxford University Press}
}

@article{brown2025constraining,
  title={Constraining primordial non-Gaussianity from the large scale structure two-point and three-point correlation functions},
  author={Brown, Z and Demina, R and Adame, AG and Avila, S and Chaussidon, E and Yuan, S and Gonzalez-Perez, V and García-Bellido, J and Levi, B and Aguilar, J and others},
  journal={Monthly Notices of the Royal Astronomical Society},
  volume={543},
  number={3},
  pages={2078--2092},
  year={2025},
  publisher={Oxford University Press}
}

@article{gil2015power,
  title={The power spectrum and bispectrum of SDSS DR11 BOSS galaxies--I. Bias and gravity},
  author={Gil-Marín, Héctor and Norena, Jorge and Verde, Licia and Percival, Will J and Wagner, Christian and Manera, Marc and Schneider, Donald P},
  journal={Monthly Notices of the Royal Astronomical Society},
  volume={451},
  number={1},
  pages={539--580},
  year={2015},
  publisher={Oxford University Press}
}

@article{gagrani2017information,
  title={Information content of the angular multipoles of redshift-space galaxy bispectrum},
  author={Gagrani, Praful and Samushia, Lado},
  journal={Monthly Notices of the Royal Astronomical Society},
  volume={467},
  number={1},
  pages={928--935},
  year={2017},
  publisher={Oxford University Press}
}

@article{slepian2025power,
  title={Power spectrum, bispectrum, 2-and 3-point correlation function, and beyond},
  author={Slepian, Zachary and Kamalinejad, Farshad and Greco, Alessandro},
  journal={arXiv preprint arXiv:2508.06762},
  year={2025}
}

@article{morawetz2025constraining,
  title={Constraining primordial non-Gaussianity with Density-Split Clustering},
  author={Morawetz, James and Paillas, Enrique and Percival, Will J},
  journal={Journal of Cosmology and Astroparticle Physics},
  volume={2025},
  number={01},
  pages={026},
  year={2025},
  publisher={IOP Publishing}
}

@article{radinovic2023euclid,
  title={Euclid: Cosmology forecasts from the void-galaxy cross-correlation function with reconstruction},
  author={Radinovi{\'c}, S and Nadathur, S and Winther, H-A and Percival, WJ and Woodfinden, A and Massara, E and Paillas, E and Contarini, S and Hamaus, N and Kovacs, A and others},
  journal={Astronomy \& Astrophysics},
  volume={677},
  pages={A78},
  year={2023},
  publisher={EDP sciences}
}

@article{sartori2025imprint,
  title={The imprint of cosmic voids from the DESI Legacy Survey DR9 Luminous Red Galaxies in the Planck 2018 lensing map through spectroscopically calibrated mocks},
  author={Sartori, S and Vielzeuf, P and Escoffier, S and Cousinou, MC and Kov{\'a}cs, A and DeRose, J and Ahlen, S and Bianchi, D and Brooks, D and Burtin, E and others},
  journal={Astronomy \& Astrophysics},
  volume={700},
  pages={A17},
  year={2025},
  publisher={EDP Sciences}
}

@article{rocher2023desi,
  title={The desi one-percent survey: exploring the halo occupation distribution of emission line galaxies with abacussummit simulations},
  author={Rocher, Antoine and Ruhlmann-Kleider, Vanina and Burtin, Etienne and Yuan, Sihan and de Mattia, Arnaud and Ross, Ashley J and Aguilar, Jessica and Ahlen, Steven and Alam, Shadab and Bianchi, Davide and others},
  journal={Journal of Cosmology and Astroparticle Physics},
  volume={2023},
  number={10},
  pages={016},
  year={2023},
  publisher={IOP Publishing}
}

@article{chudaykin2026reanalyzing,
  title={Reanalyzing DESI DR1. I. $\Lambda$ CDM constraints from the power spectrum and bispectrum},
  author={Chudaykin, Anton and Ivanov, Mikhail M and Philcox, Oliver HE},
  journal={Physical Review D},
  volume={113},
  number={6},
  pages={063502},
  year={2026},
  publisher={APS}
}

@article{mancini2024field,
  title={Field-level cosmological model selection: field-level simulation-based inference for Stage IV cosmic shear can distinguish dynamical dark energy},
  author={Mancini, A Spurio and Lin, Kiyam and McEwen, Jason D},
  journal={arXiv preprint arXiv:2410.10616},
  year={2024}
}

@article{kitanidis2021cross,
  title={Cross-correlation of Planck CMB lensing with DESI-like LRGs},
  author={Kitanidis, Ellie and White, Martin},
  journal={Monthly Notices of the Royal Astronomical Society},
  volume={501},
  number={4},
  pages={6181--6198},
  year={2021},
  publisher={Oxford University Press}
}

@article{hahn2023forward,
  title={A forward modeling approach to analyzing galaxy clustering with SIMBIG},
  author={Hahn, ChangHoon and Eickenberg, Michael and Ho, Shirley and Hou, Jiamin and Lemos, Pablo and Massara, Elena and Modi, Chirag and Moradinezhad Dizgah, Azadeh and Blancard, Bruno R{\'e}galdo-Saint and Abidi, Muntazir M},
  journal={Proceedings of the National Academy of Sciences},
  volume={120},
  number={42},
  pages={e2218810120},
  year={2023},
  publisher={National Academy of Sciences}
}

@article{lemos2024field,
  title={Field-level simulation-based inference of galaxy clustering with convolutional neural networks},
  author={Lemos, Pablo and Parker, Liam and Hahn, ChangHoon and Ho, Shirley and Eickenberg, Michael and Hou, Jiamin and Massara, Elena and Modi, Chirag and Dizgah, Azadeh Moradinezhad and Blancard, Bruno Regaldo-Saint and others},
  journal={Physical Review D},
  volume={109},
  number={8},
  pages={083536},
  year={2024},
  publisher={APS}
}

@article{bianchi2025desi_fiber_dr1,
  author  = {Bianchi, Davide and Percival, Will J. and others},
  title   = {Characterization of DESI fiber assignment incompleteness effect on 2-point clustering and mitigation methods for DR1 analysis},
  journal = {Journal of Cosmology and Astroparticle Physics},
  year    = {2025},
  volume  = {04},
  pages   = {074},
  eprint  = {2411.12025},
  archivePrefix = {arXiv},
  primaryClass  = {astro-ph.CO}
}

@article{lasker2025desi_pip,
  author  = {Lasker, Jordan and Percival, Will J. and Bianchi, Davide and others},
  title   = {Production of alternate realizations of DESI fiber assignment for unbiased clustering measurement in data and simulations},
  journal = {Journal of Cosmology and Astroparticle Physics},
  year    = {2025},
  volume  = {01},
  pages   = {127},
  doi     = {10.1088/1475-7516/2025/01/127},
  archivePrefix = {arXiv},
  primaryClass  = {astro-ph.CO}
}

@article{bianchi2018fiber_assignment,
  author  = {Bianchi, Davide and Leistedt, Boris and Sutter, P. M. and Percival, Will J.},
  title   = {Unbiased clustering estimates with the DESI fibre assignment},
  journal = {Monthly Notices of the Royal Astronomical Society},
  year    = {2018},
  volume  = {481},
  number  = {2},
  pages   = {2330--2341},
  eprint  = {1805.00951},
  archivePrefix = {arXiv},
  primaryClass  = {astro-ph.CO}
}

@article{sugiyama2019triposh,
  author  = {Sugiyama, Naonori S. and Saito, Shun and Beutler, Florian and Seo, Hee-Jong},
  title   = {A complete FFT-based decomposition formalism for the redshift-space bispectrum},
  journal = {Monthly Notices of the Royal Astronomical Society},
  year    = {2019},
  volume  = {484},
  number  = {1},
  pages   = {364--384},
  doi     = {10.1093/mnras/sty3249},
  eprint  = {1902.01066},
  archivePrefix = {arXiv},
  primaryClass  = {astro-ph.CO}
}

@article{PearsonSamushia2019,
  author  = {Pearson, David W. and Samushia, Lado},
  title   = {Computing three-point correlation function randoms counts without the randoms catalogue},
  journal = {Monthly Notices of the Royal Astronomical Society Letters},
  volume  = {486},
  number  = {1},
  pages   = {L105--L109},
  year    = {2019},
  month   = {May},
  doi     = {10.1093/mnrasl/slz072},
  url     = {https://doi.org/10.1093/mnrasl/slz072}
}

@article{hartlap2007your,
  title={Why your model parameter confidences might be too optimistic. Unbiased estimation of the inverse covariance matrix},
  author={Hartlap, J and Simon, Patrick and Schneider, P},
  journal={Astronomy \& Astrophysics},
  volume={464},
  number={1},
  pages={399--404},
  year={2007},
  publisher={EDP Sciences}
}

@ARTICLE{2019arXiv190205569A,
       author = {{Akeson}, Rachel and {Armus}, Lee and {Bachelet}, Etienne and {Bailey}, Vanessa and {Bartusek}, Lisa and {Bellini}, Andrea and {Benford}, Dominic and {Bennett}, David and {Bhattacharya}, Aparna and {Bohlin}, Ralph and {Boyer}, Martha and {Bozza}, Valerio and {Bryden}, Geoffrey and {Calchi Novati}, Sebastiano and {Carpenter}, Kenneth and {Casertano}, Stefano and {Choi}, Ami and {Content}, David and {Dayal}, Pratika and {Dressler}, Alan and {Dor{\'e}}, Olivier and {Fall}, S. Michael and {Fan}, Xiaohui and {Fang}, Xiao and {Filippenko}, Alexei and {Finkelstein}, Steven and {Foley}, Ryan and {Furlanetto}, Steven and {Kalirai}, Jason and {Gaudi}, B. Scott and {Gilbert}, Karoline and {Girard}, Julien and {Grady}, Kevin and {Greene}, Jenny and {Guhathakurta}, Puragra and {Heinrich}, Chen and {Hemmati}, Shoubaneh and {Hendel}, David and {Henderson}, Calen and {Henning}, Thomas and {Hirata}, Christopher and {Ho}, Shirley and {Huff}, Eric and {Hutter}, Anne and {Jansen}, Rolf and {Jha}, Saurabh and {Johnson}, Samson and {Jones}, David and {Kasdin}, Jeremy and {Kelly}, Patrick and {Kirshner}, Robert and {Koekemoer}, Anton and {Kruk}, Jeffrey and {Lewis}, Nikole and {Macintosh}, Bruce and {Madau}, Piero and {Malhotra}, Sangeeta and {Mandel}, Kaisey and {Massara}, Elena and {Masters}, Daniel and {McEnery}, Julie and {McQuinn}, Kristen and {Melchior}, Peter and {Melton}, Mark and {Mennesson}, Bertrand and {Peeples}, Molly and {Penny}, Matthew and {Perlmutter}, Saul and {Pisani}, Alice and {Plazas}, Andr{\'e}s and {Poleski}, Radek and {Postman}, Marc and {Ranc}, Cl{\'e}ment and {Rauscher}, Bernard and {Rest}, Armin and {Roberge}, Aki and {Robertson}, Brant and {Rodney}, Steven and {Rhoads}, James and {Rhodes}, Jason and {Ryan}, Jr., Russell and {Sahu}, Kailash and {Sand}, David and {Scolnic}, Dan and {Seth}, Anil and {Shvartzvald}, Yossi and {Siellez}, Karelle and {Smith}, Arfon and {Spergel}, David and {Stassun}, Keivan and {Street}, Rachel and {Strolger}, Louis-Gregory and {Szalay}, Alexander and {Trauger}, John and {Troxel}, M.~A. and {Turnbull}, Margaret and {van der Marel}, Roeland and {von der Linden}, Anja and {Wang}, Yun and {Weinberg}, David and {Williams}, Benjamin and {Windhorst}, Rogier and {Wollack}, Edward and {Wu}, Hao-Yi and {Yee}, Jennifer and {Zimmerman}, Neil},
        title = "{The Wide Field Infrared Survey Telescope: 100 Hubbles for the 2020s}",
      journal = {arXiv e-prints},
         year = 2019,
        month = feb,
          eid = {arXiv:1902.05569},
        pages = {arXiv:1902.05569},
          doi = {10.48550/arXiv.1902.05569},
archivePrefix = {arXiv},
       eprint = {1902.05569},
 primaryClass = {astro-ph.IM},
       adsurl = {https://ui.adsabs.harvard.edu/abs/2019arXiv190205569A}
}

@ARTICLE{2021MNRAS.507.1746E,
       author = {{Eifler}, Tim and {Miyatake}, Hironao and {Krause}, Elisabeth and {Heinrich}, Chen and {Miranda}, Vivian and {Hirata}, Christopher and {Xu}, Jiachuan and {Hemmati}, Shoubaneh and {Simet}, Melanie and {Capak}, Peter and {Choi}, Ami and {Dor{\'e}}, Olivier and {Doux}, Cyrille and {Fang}, Xiao and {Hounsell}, Rebekah and {Huff}, Eric and {Huang}, Hung-Jin and {Jarvis}, Mike and {Kruk}, Jeffrey and {Masters}, Dan and {Rozo}, Eduardo and {Scolnic}, Dan and {Spergel}, David N. and {Troxel}, Michael and {von der Linden}, Anja and {Wang}, Yun and {Weinberg}, David H. and {Wenzl}, Lukas and {Wu}, Hao-Yi},
        title = "{Cosmology with the Roman Space Telescope - multiprobe strategies}",
      journal = {\mnras},
         year = 2021,
        month = oct,
       volume = {507},
       number = {2},
        pages = {1746-1761},
          doi = {10.1093/mnras/stab1762},
archivePrefix = {arXiv},
       eprint = {2004.05271},
 primaryClass = {astro-ph.CO},
       adsurl = {https://ui.adsabs.harvard.edu/abs/2021MNRAS.507.1746E}
}

@ARTICLE{2021arXiv210912216Z,
       author = {{Zhai}, Zhongxu and {Wang}, Yun and {Benson}, Andrew and {Colbert}, James and {Bagley}, Micaela and {Henry}, Alaina and {Baronchelli}, Ivano},
        title = "{Simulating properties of emission line galaxies from Nancy Grace Roman Space Telescope}",
      journal = {arXiv e-prints},
         year = 2021,
        month = sep,
          eid = {arXiv:2109.12216},
        pages = {arXiv:2109.12216},
          doi = {10.48550/arXiv.2109.12216},
archivePrefix = {arXiv},
       eprint = {2109.12216},
 primaryClass = {astro-ph.GA},
       adsurl = {https://ui.adsabs.harvard.edu/abs/2021arXiv210912216Z}
}

@ARTICLE{2021MNRAS.501.3490Z,
       author = {{Zhai}, Zhongxu and {Chuang}, Chia-Hsun and {Wang}, Yun and {Benson}, Andrew and {Yepes}, Gustavo},
        title = "{Clustering in the simulated H {\ensuremath{\alpha}} galaxy redshift survey from Nancy Grace Roman Space Telescope}",
      journal = {\mnras},
         year = 2021,
        month = mar,
       volume = {501},
       number = {3},
        pages = {3490-3501},
          doi = {10.1093/mnras/staa3911},
archivePrefix = {arXiv},
       eprint = {2008.09746},
 primaryClass = {astro-ph.CO},
       adsurl = {https://ui.adsabs.harvard.edu/abs/2021MNRAS.501.3490Z}
}

@ARTICLE{2019MNRAS.490.3667Z,
       author = {{Zhai}, Zhongxu and {Benson}, Andrew and {Wang}, Yun and {Yepes}, Gustavo and {Chuang}, Chia-Hsun},
        title = "{Prediction of H {\ensuremath{\alpha}} and [O III] emission line galaxy number counts for future galaxy redshift surveys}",
      journal = {\mnras},
         year = 2019,
        month = dec,
       volume = {490},
       number = {3},
        pages = {3667-3678},
          doi = {10.1093/mnras/stz2844},
archivePrefix = {arXiv},
       eprint = {1907.09680},
 primaryClass = {astro-ph.GA},
       adsurl = {https://ui.adsabs.harvard.edu/abs/2019MNRAS.490.3667Z}
}

@ARTICLE{2022ApJ...928....1W,
       author = {{Wang}, Yun and {Zhai}, Zhongxu and {Alavi}, Anahita and {Massara}, Elena and {Pisani}, Alice and {Benson}, Andrew and {Hirata}, Christopher M. and {Samushia}, Lado and {Weinberg}, David H. and {Colbert}, James and {Dor{\'e}}, Olivier and {Eifler}, Tim and {Heinrich}, Chen and {Ho}, Shirley and {Krause}, Elisabeth and {Padmanabhan}, Nikhil and {Spergel}, David and {Teplitz}, Harry I.},
        title = "{The High Latitude Spectroscopic Survey on the Nancy Grace Roman Space Telescope}",
      journal = {\apj},
         year = 2022,
        month = mar,
       volume = {928},
       number = {1},
          eid = {1},
        pages = {1},
          doi = {10.3847/1538-4357/ac4973},
archivePrefix = {arXiv},
       eprint = {2110.01829},
 primaryClass = {astro-ph.CO},
       adsurl = {https://ui.adsabs.harvard.edu/abs/2022ApJ...928....1W}
}

@misc{novellmasot2025fullshape,
      title={Full-Shape analysis of the power spectrum and bispectrum of DESI DR1 LRG and QSO samples}, 
      author={S. Novell-Masot and H. Gil-Marín and L. Verde and J. Aguilar and S. Ahlen and S. Bailey and S. BenZvi and D. Bianchi and D. Brooks and E. Buckley-Geer and A. Carnero Rosell and E. Chaussidon and T. Claybaugh and S. Cole and A. Cuceu and K. S. Dawson and A. de la Macorra and R. Demina and A. Dey and B. Dey and P. Doel and S. Ferraro and A. Font-Ribera and J. E. Forero-Romero and E. Gaztañaga and S. Gontcho A Gontcho and A. X. Gonzalez-Morales and G. Gutierrez and H. K. Herrera-Alcantar and K. Honscheid and C. Howlett and S. Juneau and R. Kehoe and D. Kirkby and T. Kisner and A. Kremin and C. Lamman and M. Landriau and L. Le Guillou and M. E. Levi and C. Magneville and M. Manera and A. Meisner and R. Miquel and J. Moustakas and A. Muñoz-Gutiérrez and A. D. Myers and S. Nadathur and G. Niz and H. E. Noriega and W. J. Percival and C. Poppett and F. Prada and I. Pérez-Ràfols and A. J. Ross and G. Rossi and L. Samushia and E. Sanchez and D. Schlegel and M. Schubnell and H. Seo and J. Silber and D. Sprayberry and G. Tarlé and M. Vargas-Magaña and B. A. Weaver and P. Zarrouk and R. Zhou and H. Zou},
      year={2025},
      eprint={2503.09714},
      archivePrefix={arXiv},
      primaryClass={astro-ph.CO},
      url={https://arxiv.org/abs/2503.09714}, 
}

@misc{novellmasot2026cosmological,
      title={Cosmological constraints from the DESI DR1 joint power spectrum and bispectrum analysis}, 
      author={S. Novell-Masot and H. Gil-Marín and L. Verde and J. Aguilar and S. Ahlen and D. Bianchi and D. Brooks and F. J. Castander and T. Claybaugh and S. Cole and A. de la Macorra and J. Della Costa and S. Ferraro and A. Font-Ribera and J. E. Forero-Romero and E. Gaztañaga and S. Gontcho A Gontcho and A. X. Gonzalez-Morales and G. Gutierrez and J. Guy and C. Hahn and H. K. Herrera-Alcantar and K. Honscheid and C. Howlett and M. Ishak and J. Jimenez and R. Joyce and R. Kehoe and D. Kirkby and A. Kremin and C. Lamman and L. Le Guillou and M. Manera and A. Meisner and R. Miquel and S. Nadathur and G. Niz and W. J. Percival and I. Pérez-Ràfols and G. Rossi and L. Samushia and E. Sanchez and E. F. Schlafly and D. Schlegel and M. Schubnell and J. Silber and D. Sprayberry and G. Tarlé and B. A. Weaver and C. Zhao and R. Zhou},
      year={2026},
      eprint={2603.19356},
      archivePrefix={arXiv},
      primaryClass={astro-ph.CO},
      url={https://arxiv.org/abs/2603.19356}, 
}

@article{sugiyama2023new,
   title={New constraints on cosmological modified gravity theories from anisotropic three-point correlation functions of BOSS DR12 galaxies},
   volume={523},
   ISSN={1365-2966},
   url={http://dx.doi.org/10.1093/mnras/stad1505},
   DOI={10.1093/mnras/stad1505},
   number={2},
   journal={Monthly Notices of the Royal Astronomical Society},
   publisher={Oxford University Press (OUP)},
   author={Sugiyama, Naonori S and Yamauchi, Daisuke and Kobayashi, Tsutomu and Fujita, Tomohiro and Arai, Shun and Hirano, Shin’ichi and Saito, Shun and Beutler, Florian and Seo, Hee-Jong},
   year={2023},
   month=May, pages={3133–3191} }

@article{farina2024modeling,
   title={Modeling and measuring the anisotropic halo 3-point correlation function: a coordinated study},
   volume={2026},
   ISSN={1475-7516},
   url={http://dx.doi.org/10.1088/1475-7516/2026/02/028},
   DOI={10.1088/1475-7516/2026/02/028},
   number={02},
   journal={Journal of Cosmology and Astroparticle Physics},
   publisher={IOP Publishing},
   author={Farina, A. and Veropalumbo, A. and Branchini, E. and Guidi, M.},
   year={2026},
   month=Feb, pages={028} }

@article{Yuan_2022,
   title={Stringent σ8 constraints from small-scale galaxy clustering using a hybrid MCMC + emulator framework},
   volume={515},
   ISSN={1365-2966},
   url={http://dx.doi.org/10.1093/mnras/stac1830},
   DOI={10.1093/mnras/stac1830},
   number={1},
   journal={Monthly Notices of the Royal Astronomical Society},
   publisher={Oxford University Press (OUP)},
   author={Yuan, Sihan and Garrison, Lehman H and Eisenstein, Daniel J and Wechsler, Risa H},
   year={2022},
   month=July, pages={871–896} }

@article{Yuan_2023,
   title={Full forward model of galaxy clustering statistics with <scp>AbacusSummit</scp> light cones},
   volume={520},
   ISSN={1365-2966},
   url={http://dx.doi.org/10.1093/mnras/stad550},
   DOI={10.1093/mnras/stad550},
   number={4},
   journal={Monthly Notices of the Royal Astronomical Society},
   publisher={Oxford University Press (OUP)},
   author={Yuan, Sihan and Hadzhiyska, Boryana and Abel, Tom},
   year={2023},
   month=Feb, pages={6283–6298} }

@misc{cuestalazaro2023sunbirdsimulationbasedmodelfullshape,
      title={SUNBIRD: A simulation-based model for full-shape density-split clustering}, 
      author={Carolina Cuesta-Lazaro and Enrique Paillas and Sihan Yuan and Yan-Chuan Cai and Seshadri Nadathur and Will J. Percival and Florian Beutler and Arnaud de Mattia and Daniel Eisenstein and Daniel Forero-Sanchez and Nelson Padilla and Mathilde Pinon and Vanina Ruhlmann-Kleider and Ariel G. Sánchez and Georgios Valogiannis and Pauline Zarrouk},
      year={2023},
      eprint={2309.16539},
      archivePrefix={arXiv},
      primaryClass={astro-ph.CO},
      url={https://arxiv.org/abs/2309.16539}, 
}

@article{Valogiannis_2024,
   title={Precise cosmological constraints from BOSS galaxy clustering with a simulation-based emulator of the wavelet scattering transform},
   volume={109},
   ISSN={2470-0029},
   url={http://dx.doi.org/10.1103/PhysRevD.109.103503},
   DOI={10.1103/physrevd.109.103503},
   number={10},
   journal={Physical Review D},
   publisher={American Physical Society (APS)},
   author={Valogiannis, Georgios and Yuan, Sihan and Dvorkin, Cora},
   year={2024},
   month=May }

@misc{lange2025cosmologicalconstraintsfullscaleclustering,
      title={Cosmological Constraints from Full-Scale Clustering and Galaxy-Galaxy Lensing with DESI DR1}, 
      author={Johannes U. Lange and Alexandra Wells and Andrew Hearin and Gillian Beltz-Mohrmann and Alexie Leauthaud and Sven Heydenreich and Chris Blake and Jessica Nicole Aguilar and Steven Ahlen and Abhijeet Anand and Davide Bianchi and David Brooks and Francisco Javier Castander and Todd Claybaugh and Shaun Cole and Andrei Cuceu and Kyle Dawson and Axel de la Macorra and Biprateep Dey and Peter Doel and Ann Elliott and Ni Putu Audita Placida Emas and Simone Ferraro and Andreu Font-Ribera and Jaime E. Forero-Romero and Cristhian Garcia-Quintero and Enrique Gaztañaga and Satya Gontcho A Gontcho and Gaston Gutierrez and Julien Guy and Klaus Honscheid and Dragan Huterer and Mustapha Ishak and Shahab Joudaki and Dick Joyce and Robert Kehoe and David Kirkby and Theodore Kisner and Anthony Kremin and Alex Krolewski and Ofer Lahav and Claire Lamman and Martin Landriau and Laurent Le Guillou and Michael Levi and Marc Manera and Paul Martini and Aaron Meisner and Ramon Miquel and John Moustakas and Eva-Maria Mueller and Seshadri Nadathur and Jeffrey A. Newman and Gustavo Niz and Nathalie Palanque-Delabrouille and Will Percival and Claire Poppett and Anna Porredon and Francisco Prada and Ignasi Pérez-Ràfols and Amy Robertson and Graziano Rossi and Rossana Ruggeri and Eusebio Sanchez and Christoph Saulder and David Schlegel and Michael Schubnell and Agne Semenaite and Hee-Jong Seo and Joseph Harry Silber and David Sprayberry and Zechang Sun and Gregory Tarlé and Mariana Vargas Magana and Benjamin Alan Weaver and Risa Wechsler and Pauline Zarrouk and Rongpu Zhou and Hu Zou},
      year={2025},
      eprint={2512.15962},
      archivePrefix={arXiv},
      primaryClass={astro-ph.CO},
      url={https://arxiv.org/abs/2512.15962}, 
}

@misc{dumerchat2026emulatinggalaxypeculiarvelocity,
      title={Emulating galaxy and peculiar velocity clustering on non-linear scales}, 
      author={T. Dumerchat and J. Bautista and C. Ravoux and J. Aguilar and S. Ahlen and S. BenZvi and D. Bianchi and D. Brooks and T. Claybaugh and A. de la Macorra and P. Doel and S. Ferraro and J. E. Forero-Romero and E. Gaztañaga and S. Gontcho A Gontcho and G. Gutierrez and C. Hahn and C. Howlett and M. Ishak and R. Joyce and D. Kirkby and A. Kremin and C. Lamman and M. Landriau and L. Le Guillou and M. Manera and R. Miquel and S. Nadathur and W. J. Percival and F. Prada and I. Pérez-Ràfols and G. Rossi and E. Sanchez and D. Schlegel and M. Schubnell and J. Silber and D. Sprayberry and G. Tarlé and B. A. Weaver and H. Zou},
      year={2026},
      eprint={2602.03382},
      archivePrefix={arXiv},
      primaryClass={astro-ph.CO},
      url={https://arxiv.org/abs/2602.03382}, 
}

@misc{kamalinejad2026detectionbaryonacousticoscillation,
      title={First Detection of the Baryon Acoustic Oscillation (BAO) Feature in the 3-Point Correlation Function of DESI DR1 Luminous Red Galaxies}, 
      author={Farshad Kamalinejad and Zachary Slepian and Alex Krolewski and Alessandro Greco and William Ortolá Leonard and Jessica Chellino and Matthew Reinhard and Elena Fernández-García and Francisco Prada and J. Aguilar and S. Ahlen and A. Anand and C. Bebek and D. Bianchi and D. Brooks and T. Claybaugh and A. Cuceu and K. S. Dawson and A. de la Macorra and R. Demina and P. Doel and J. Edelstein and J. E. Forero-Romero and E. Gaztañaga and S. Gontcho A Gontcho and G. Gutierrez and H. K. Herrera-Alcantar and K. Honscheid and C. Howlett and D. Huterer and M. Ishak and R. Joyce and S. Juneau and D. Kirkby and T. Kisner and A. Kremin and O. Lahav and C. Lamman and M. Landriau and L. Le Guillou and M. Manera and A. Meisner and R. Miquel and J. A. Newman and W. J. Percival and C. Poppett and I. Pérez-Ràfols and L. Samushia and E. Sanchez and D. Schlegel and M. Schubnell and H. Seo and J. Silber and D. Sprayberry and G. Tarlé and B. A. Weaver and C. Zhao and H. Zou},
      year={2026},
      eprint={2602.16134},
      archivePrefix={arXiv},
      primaryClass={astro-ph.CO},
      url={https://arxiv.org/abs/2602.16134}, 
}

@article{Burger_2024,
   title={Cosmological parameters from the joint analysis of density split and second order statistics: An emulator based on the halo occupation distribution},
   volume={110},
   ISSN={2470-0029},
   url={http://dx.doi.org/10.1103/PhysRevD.110.083517},
   DOI={10.1103/physrevd.110.083517},
   number={8},
   journal={Physical Review D},
   publisher={American Physical Society (APS)},
   author={Burger, Pierre A. and Paillas, Enrique and Hudson, Michael J.},
   year={2024},
   month=Oct }

@article{Slepian_2017,
   title={Detection of baryon acoustic oscillation features in the large-scale three-point correlation function of SDSS BOSS DR12 CMASS galaxies},
   volume={469},
   ISSN={1365-2966},
   url={http://dx.doi.org/10.1093/mnras/stx488},
   DOI={10.1093/mnras/stx488},
   number={2},
   journal={Monthly Notices of the Royal Astronomical Society},
   publisher={Oxford University Press (OUP)},
   author={Slepian, Zachary and Eisenstein, Daniel J. and Brownstein, Joel R. and Chuang, Chia-Hsun and Gil-Marín, Héctor and Ho, Shirley and Kitaura, Francisco-Shu and Percival, Will J. and Ross, Ashley J. and Rossi, Graziano and Seo, Hee-Jong and Slosar, Anže and Vargas-Magaña, Mariana},
   year={2017},
   month=Mar, pages={1738–1751} }

@ARTICLE{2021MNRAS.505..628S,
       author = {{Samushia}, Lado and {Slepian}, Zachary and {Villaescusa-Navarro}, Francisco},
        title = "{Information content of higher order galaxy correlation functions}",
      journal = {\mnras},
         year = 2021,
        month = jul,
       volume = {505},
       number = {1},
        pages = {628-641},
          doi = {10.1093/mnras/stab1199},
archivePrefix = {arXiv},
       eprint = {2102.01696},
 primaryClass = {astro-ph.CO},
       adsurl = {https://ui.adsabs.harvard.edu/abs/2021MNRAS.505..628S}
}

@ARTICLE{2018MNRAS.478.4500P,
       author = {{Pearson}, David W. and {Samushia}, Lado},
        title = "{A Detection of the Baryon Acoustic Oscillation features in the SDSS BOSS DR12 Galaxy Bispectrum}",
      journal = {\mnras},
         year = 2018,
        month = aug,
       volume = {478},
       number = {4},
        pages = {4500-4512},
          doi = {10.1093/mnras/sty1266},
archivePrefix = {arXiv},
       eprint = {1712.04970},
 primaryClass = {astro-ph.CO},
       adsurl = {https://ui.adsabs.harvard.edu/abs/2018MNRAS.478.4500P}
}

@ARTICLE{2024MNRAS.531.3326B,
       author = {{Behera}, Jayashree and {Rezaie}, Mehdi and {Samushia}, Lado and {Ereza}, Julia},
        title = "{Modelling the BAO feature in bispectrum}",
      journal = {\mnras},
         year = 2024,
        month = jul,
       volume = {531},
       number = {3},
        pages = {3326-3335},
          doi = {10.1093/mnras/stae1161},
archivePrefix = {arXiv},
       eprint = {2312.05942},
 primaryClass = {astro-ph.CO},
       adsurl = {https://ui.adsabs.harvard.edu/abs/2024MNRAS.531.3326B}
}

@ARTICLE{2020A&A...641A...6P,
       author = {{Planck Collaboration} and {Aghanim}, N. and {Akrami}, Y. and {Ashdown}, M. and {Aumont}, J. and {Baccigalupi}, C. and {Ballardini}, M. and {Banday}, A.~J. and {Barreiro}, R.~B. and {Bartolo}, N. and {Basak}, S. and {Battye}, R. and {Benabed}, K. and {Bernard}, J.-P. and {Bersanelli}, M. and {Bielewicz}, P. and {Bock}, J.~J. and {Bond}, J.~R. and {Borrill}, J. and {Bouchet}, F.~R. and {Boulanger}, F. and {Bucher}, M. and {Burigana}, C. and {Butler}, R.~C. and {Calabrese}, E. and {Cardoso}, J.-F. and {Carron}, J. and {Challinor}, A. and {Chiang}, H.~C. and {Chluba}, J. and {Colombo}, L.~P.~L. and {Combet}, C. and {Contreras}, D. and {Crill}, B.~P. and {Cuttaia}, F. and {de Bernardis}, P. and {de Zotti}, G. and {Delabrouille}, J. and {Delouis}, J.-M. and {Di Valentino}, E. and {Diego}, J.~M. and {Dor{\'e}}, O. and {Douspis}, M. and {Ducout}, A. and {Dupac}, X. and {Dusini}, S. and {Efstathiou}, G. and {Elsner}, F. and {En{\ss}lin}, T.~A. and {Eriksen}, H.~K. and {Fantaye}, Y. and {Farhang}, M. and {Fergusson}, J. and {Fernandez-Cobos}, R. and {Finelli}, F. and {Forastieri}, F. and {Frailis}, M. and {Fraisse}, A.~A. and {Franceschi}, E. and {Frolov}, A. and {Galeotta}, S. and {Galli}, S. and {Ganga}, K. and {G{\'e}nova-Santos}, R.~T. and {Gerbino}, M. and {Ghosh}, T. and {Gonz{\'a}lez-Nuevo}, J. and {G{\'o}rski}, K.~M. and {Gratton}, S. and {Gruppuso}, A. and {Gudmundsson}, J.~E. and {Hamann}, J. and {Handley}, W. and {Hansen}, F.~K. and {Herranz}, D. and {Hildebrandt}, S.~R. and {Hivon}, E. and {Huang}, Z. and {Jaffe}, A.~H. and {Jones}, W.~C. and {Karakci}, A. and {Keih{\"a}nen}, E. and {Keskitalo}, R. and {Kiiveri}, K. and {Kim}, J. and {Kisner}, T.~S. and {Knox}, L. and {Krachmalnicoff}, N. and {Kunz}, M. and {Kurki-Suonio}, H. and {Lagache}, G. and {Lamarre}, J.-M. and {Lasenby}, A. and {Lattanzi}, M. and {Lawrence}, C.~R. and {Le Jeune}, M. and {Lemos}, P. and {Lesgourgues}, J. and {Levrier}, F. and {Lewis}, A. and {Liguori}, M. and {Lilje}, P.~B. and {Lilley}, M. and {Lindholm}, V. and {L{\'o}pez-Caniego}, M. and {Lubin}, P.~M. and {Ma}, Y.-Z. and {Mac{\'\i}as-P{\'e}rez}, J.~F. and {Maggio}, G. and {Maino}, D. and {Mandolesi}, N. and {Mangilli}, A. and {Marcos-Caballero}, A. and {Maris}, M. and {Martin}, P.~G. and {Martinelli}, M. and {Mart{\'\i}nez-Gonz{\'a}lez}, E. and {Matarrese}, S. and {Mauri}, N. and {McEwen}, J.~D. and {Meinhold}, P.~R. and {Melchiorri}, A. and {Mennella}, A. and {Migliaccio}, M. and {Millea}, M. and {Mitra}, S. and {Miville-Desch{\^e}nes}, M.-A. and {Molinari}, D. and {Montier}, L. and {Morgante}, G. and {Moss}, A. and {Natoli}, P. and {N{\o}rgaard-Nielsen}, H.~U. and {Pagano}, L. and {Paoletti}, D. and {Partridge}, B. and {Patanchon}, G. and {Peiris}, H.~V. and {Perrotta}, F. and {Pettorino}, V. and {Piacentini}, F. and {Polastri}, L. and {Polenta}, G. and {Puget}, J.-L. and {Rachen}, J.~P. and {Reinecke}, M. and {Remazeilles}, M. and {Renzi}, A. and {Rocha}, G. and {Rosset}, C. and {Roudier}, G. and {Rubi{\~n}o-Mart{\'\i}n}, J.~A. and {Ruiz-Granados}, B. and {Salvati}, L. and {Sandri}, M. and {Savelainen}, M. and {Scott}, D. and {Shellard}, E.~P.~S. and {Sirignano}, C. and {Sirri}, G. and {Spencer}, L.~D. and {Sunyaev}, R. and {Suur-Uski}, A.-S. and {Tauber}, J.~A. and {Tavagnacco}, D. and {Tenti}, M. and {Toffolatti}, L. and {Tomasi}, M. and {Trombetti}, T. and {Valenziano}, L. and {Valiviita}, J. and {Van Tent}, B. and {Vibert}, L. and {Vielva}, P. and {Villa}, F. and {Vittorio}, N. and {Wandelt}, B.~D. and {Wehus}, I.~K. and {White}, M. and {White}, S.~D.~M. and {Zacchei}, A. and {Zonca}, A.},
        title = "{Planck 2018 results. VI. Cosmological parameters}",
      journal = {\aap},
         year = 2020,
        month = sep,
       volume = {641},
          eid = {A6},
        pages = {A6},
          doi = {10.1051/0004-6361/201833910},
archivePrefix = {arXiv},
       eprint = {1807.06209},
 primaryClass = {astro-ph.CO},
       adsurl = {https://ui.adsabs.harvard.edu/abs/2020A&A...641A...6P}
}

@misc{spezzati2026forecastingneutrinomassconstraints,
      title={Forecasting neutrino mass constraints from the Nancy Grace Roman Space Telescope}, 
      author={Francesco Spezzati and Yun Wang and Andrew Hearin},
      year={2026},
      eprint={2604.14504},
      archivePrefix={arXiv},
      primaryClass={astro-ph.CO},
      url={https://arxiv.org/abs/2604.14504}, 
}

@misc{yuan2023desionepercentsurveyexploring,
      title={The DESI One-Percent Survey: Exploring the Halo Occupation Distribution of Luminous Red Galaxies and Quasi-Stellar Objects with AbacusSummit}, 
      author={Sihan Yuan and Hanyu Zhang and Ashley J. Ross and Jamie Donald-McCann and Boryana Hadzhiyska and Risa H. Wechsler and Zheng Zheng and Shadab Alam and Violeta Gonzalez-Perez and Jessica Nicole Aguilar and Steven Ahlen and Davide Bianchi and David Brooks and Axel de la Macorra and Kevin Fanning and Jaime E. Forero-Romero and Klaus Honscheid and Mustapha Ishak and Robert Kehoe and James Lasker and Martin Landriau and Marc Manera and Paul Martini and Aaron Meisner and Ramon Miquel and John Moustakas and Seshadri Nadathur and Jeffrey A. Newman and Jundan Nie and Will Percival and Claire Poppett and Antoine Rocher and Graziano Rossi and Eusebio Sanchez and Lado Samushia and Michael Schubnell and Hee-Jong Seo and Gregory Tarle and Benjamin Alan Weaver and Jiaxi Yu and Zhimin Zhou and Hu Zou},
      year={2023},
      eprint={2306.06314},
      archivePrefix={arXiv},
      primaryClass={astro-ph.CO},
      url={https://arxiv.org/abs/2306.06314}, 
}

@article{Mena_Fern_ndez_2025,
   title={HOD-dependent systematics for luminous red galaxies in the DESI 2024 BAO analysis},
   volume={2025},
   ISSN={1475-7516},
   url={http://dx.doi.org/10.1088/1475-7516/2025/01/133},
   DOI={10.1088/1475-7516/2025/01/133},
   number={01},
   journal={Journal of Cosmology and Astroparticle Physics},
   publisher={IOP Publishing},
   author={Mena-Fernández, J. and Garcia-Quintero, C. and Yuan, S. and Hadzhiyska, B. and Alves, O. and Rashkovetskyi, M. and Seo, H. and Padmanabhan, N. and Nadathur, S. and Howlett, C. and Alam, S. and Rocher, A. and Ross, A.J. and Sanchez, E. and Ishak, M. and Aguilar, J. and Ahlen, S. and Andrade, U. and BenZvi, S. and Brooks, D. and Burtin, E. and Chen, S. and Chen, X. and Claybaugh, T. and Cole, S. and de la Macorra, A. and de Mattia, A. and Dey, Arjun and Dey, B. and Ding, Z. and Doel, P. and Fanning, K. and Forero-Romero, J.E. and Gaztañaga, E. and Gil-Marín, H. and Gontcho, S.Gontcho A. and Gutierrez, G. and Guy, J. and Hahn, C. and Honscheid, K. and Juneau, S. and Kremin, A. and Landriau, M. and Le Guillou, L. and Levi, M.E. and Manera, M. and Martini, P. and Medina-Varela, L. and Meisner, A. and Miquel, R. and Moustakas, J. and Mueller, E. and Muñoz-Gutiérrez, A. and Myers, A.D. and Newman, J.A. and Nie, J. and Niz, G. and Paillas, E. and Palanque-Delabrouille, N. and Percival, W.J. and Poppett, C. and Pérez-Fernández, A. and Rosado-Marin, A. and Rossi, G. and Ruggeri, R. and Saulder, C. and Schlegel, D. and Schubnell, M. and Sprayberry, D. and Tarlé, G. and Vargas-Magaña, M. and Weaver, B.A. and Yu, J. and Zhang, H. and Zou, H.},
   year={2025},
   month=Jan, pages={133} }

@article{foreman2013emcee,
  title={emcee: the MCMC hammer},
  author={Foreman-Mackey, Daniel and Hogg, David W and Lang, Dustin and Goodman, Jonathan},
  journal={Publications of the Astronomical Society of the Pacific},
  volume={125},
  number={925},
  pages={306--312},
  year={2013},
  publisher={University of Chicago Press}
}

@article{novell2024approximations,
  title={On approximations of the redshift-space bispectrum and power spectrum multipoles covariance matrix},
  author={Novell-Masot, Sergi and Gil-Mar{\'\i}n, H{\'e}ctor and Verde, Licia},
  journal={Journal of Cosmology and Astroparticle Physics},
  volume={2024},
  number={06},
  pages={048},
  year={2024},
  publisher={IOP Publishing}
}

@article{umeh2021optimal,
  title={Optimal computation of anisotropic galaxy three point correlation function multipoles using 2DFFTLOG formalism},
  author={Umeh, Obinna},
  journal={Journal of Cosmology and Astroparticle Physics},
  volume={2021},
  number={05},
  pages={035},
  year={2021},
  publisher={IOP Publishing}
}

@article{bocquet2016pygtc,
  title={pygtc: beautiful parameter covariance plots (aka. Giant Triangle Confusograms)},
  author={Bocquet, Sebastian and Carter, Faustin W},
  journal={Journal of Open Source Software},
  volume={1},
  number={6},
  pages={46},
  year={2016}
}




\end{document}